\documentclass[apj]{emulateapj}
\shortauthors{D. Saylor, et al.}
\slugcomment{Accepted for Publication in AJ}

\usepackage{subfigure}
\usepackage{float}
\usepackage{amsmath}
\usepackage{graphicx}
\usepackage{morefloats}
\usepackage{relsize}

\begin{document}

\title{Light Curve Modulation of Low Mass Stars in \emph{K2}. I. \\
    Identification of 508 Fast Rotators in the Solar Neighborhood}
\author{Dicy Saylor\altaffilmark{1}, Sebastien Lepine\altaffilmark{1}, Ian Crossfield\altaffilmark{2},\altaffilmark{3}, and Erik A. Petigura\altaffilmark{4}}
\altaffiltext{1}{Department of Physics and Astronomy, Georgia State University, 25 Park Place South, Suite 605, Atlanta GA, 30303, USA}
\altaffiltext{2}{Lunar \& Planetary Laboratory, University of Arizona, 1629
E. University Blvd., Tucson, AZ, USA}
\altaffiltext{3}{ Department of Physics, Massachusetts Institute of Technology, Cambridge, MA, USA}
\altaffiltext{4}{California Institute of Technology, Pasadena, CA, USA; Hubble Fellow}
\shorttitle{\emph{K2} Fast Rotators}

\begin{abstract}
The \emph{K2} mission is targeting large numbers of nearby ($d<100$ pc) GKM dwarfs selected from the SUPERBLINK proper motion survey ($\mu>40$ mas yr$^{-1}$, $V<20$). Additionally, the mission is targeting low-mass, high proper motion stars associated with the local ($d<500$ pc) Galactic halo population also selected from SUPERBLINK. \emph{K2} campaigns 0 through 8 monitored a total of 27,382 of these cool main-sequence stars. We used the auto-correlation function to search for fast rotators by identifying short-period photometric modulations in the \emph{K2} light curves. We identified 508 candidate fast rotators with rotation periods $<4$ days that show light curve modulations consistent with star spots. Their kinematics show low average transverse velocities, suggesting they are part of the young disk population. A subset (13) of the fast rotators are found among those targets with colors and kinematics consistent with the local Galactic halo population and may represent stars spun up by tidal interactions in close binary systems. We further demonstrate the M dwarf fast rotators selected from the \emph{K2} light curves are significantly more likely to have UV excess, and discuss the potential of the \emph{K2} mission to identify new nearby young GKM dwarfs on the basis of their fast rotation rates. Finally, we discuss the possible use of local halo stars as fiducial, non-variable sources in the \emph{Kepler} fields.  

\end{abstract}

\subjectheadings{lstars: rotation, stars: activity, stars: low-mass, stars: Population II, surveys}

\section{Introduction}
Stars with convective interiors like the Sun have active magnetic fields that produce relatively cool spots on their surface, which appear darker than the surrounding photosphere. As the star rotates, the disk-integrated brightness of the star changes as spots come in and out of view. Using time series analysis, we can determine the star's rotation rate from the modulation of the light curve over time \citep{McQ13}. In recent years, a vast amount of high precision photometric data from \emph{Kepler} and \emph{K2} has facilitated the search for exoplanets. An additional benefit of these high-precision, long term data sets is the ability to determine rotation rates from photometric modulation due to star spots \citep{Rap14}. \\ \indent
A star's rotation rate is generally well-correlated with its age, a phenomenon that has been empirically demonstrated since the 1970s \citep{Sku72,Bar07,MH08}. The rotation-age relationship, also known as \emph{gyrochronolgy} appears to be relatively reliable for Solar-type stars, though it possibly breaks down in the low-mass regime due to changes in the rotational braking mechanism produced by magnetic fields \citep{Reiners08}. However, the rule that fast rotation indicates relative youth (i.e. stellar ages $<1$ Gyr) still holds true in general for early to mid-type M dwarfs \citep{N15}. \\ \indent
The identification of nearby young M dwarfs is an outstanding problem in stellar astronomy. Nearby ($d<100$ pc) young stars are typically identified by moving group membership; proper motion and radial velocity measurements are used to associate a low-mass dwarf with known groups of young (usually more massive) stars. Kinematic data can indicate youth in two ways. In the more general case, where stellar radial velocities are unavailable, proper motions combined with photometric distances may suggest that a star is ``kinematically young'' by placing it among the young disk population ($<2-5$ Gyr). On the other hand, if a star has proper motion, distance, and radial velocity data available, we can calculate UVW space velocities and determine group membership \citep{Gagne14}. Identification of group membership generally indicates a much younger age ($<100$ Myr) as moving groups tend to disperse on relatively short timescale. \\ \indent
New members to moving groups are determined by combining UVW velocities with a secondary youth indicator, such as strong X-ray flux, UV excess, or H$\alpha$ emission. Young M dwarfs have strong magnetic fields making them chromospherically active and producing bright X-ray emission \citep{Flem95}. Many nearby, young M dwarfs are indeed detected as \emph{ROSAT} sources. Active M dwarfs can also show strong UV emission \citep{A2015}, and many are detected in the \emph{GALEX} survey.  If a cool M dwarf is detected in the X-ray or UV, it can safely be designated as young \citep{Evgenya12}. These data are difficult to obtain for most M dwarfs due to their low luminosities at all wavelengths, unless they are relatively nearby ($d<100$pc). This has led to the suggestion that many M dwarfs may be missing from surveys of nearby moving group members because their X-ray or UV brightness is below survey detection limits \citep{Schlieder11}.\\ \indent

\begin{figure*}[]
\centering
  \includegraphics[width=160mm,height=140mm]{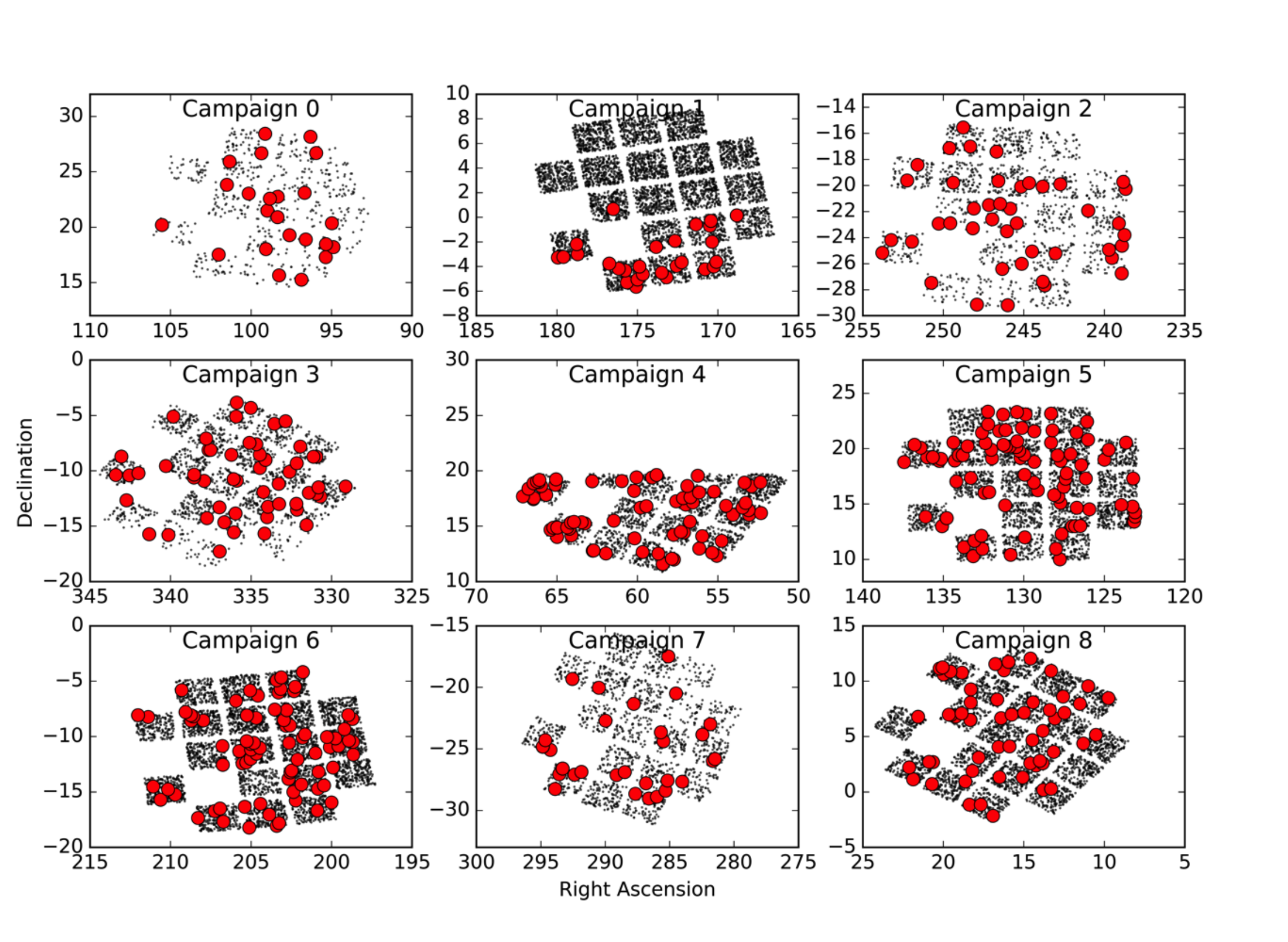}
\caption{Positions of SBK2 targets in campaigns 0-8. The black points represent all SBK2 targets, the red circles represent the fast rotators $(P_{rot}$ $\leq$ $4$ days) identified in our search.}
\label{fig:position}
\end{figure*}

The current method for identifying and confirming young, nearby stars requires that a candidate has both a proper motion consistent with any one of the known moving groups in addition to having detected flux in the X-ray and/or UV. The standard procedure then requires one to obtain a spectrum of the candidate and identify other activity features such as a strong $H{\alpha}$ in emission. However, stars that do not meet both criteria (kinematics $+$ X-ray/UV detection) are generally not considered to show enough evidence of youth, and are typically not considered for follow-up spectroscopic observations. \\ \indent
Lists of confirmed moving group members have fewer M dwarfs than one would expect, compared with the large numbers of M dwarfs normally found among field stars. A possible explanation for the small number of M dwarfs is that we fail to identify them because our criteria are too strict. Perhaps some young M dwarfs have X-ray or UV emission that is not bright enough to be detected in current surveys; for example the limiting magnitude of \emph{GALEX} is $NUV$ $\sim$ $20.5$. There could be an advantage in finding an alternative or additional diagnostic for young stars, e.g. fast rotation. We investigate the potential for using rapid rotation as a youth indicator in the identification of nearby young moving group members based on the hypothesis that young M dwarfs should have rapid rotation even if no clear chromospheric activity (X-ray, UV excess) is detected. With the photometric data available from \emph{K2} we have the potential to determine youth via rotation for hundreds of M dwarfs. \\ \indent
In this paper, we present our data selection and reduction in Section \ref{sec:data}, data analysis in Section \ref{sec:acf}, and a discussion of our results in Sections \ref{sec:selection}, \ref{sec:kinematics}, and \ref{sec:uvexcess}. We present our conclusions in Section \ref{sec:conclusions}. 

\section{Data Acquisition \& Reduction}
\label{sec:data}
\label{dataselect}
The current version of the SUPERBLINK proper motion catalog lists stars with proper motions $\mu>40$ mas yr$^{-1}$ and visual magnitudes $V$ $<$ $20$ over the entire sky north of $Decl.=-30$; this area includes all fields observable in \emph{K2} \citep{Lepine05,Lep11}. Lists of nearby stars with estimated distances $d<100$pc were assembled from this catalog, and proposed as targets of interest in the first calls for \emph{K2} targets. Several thousand SUPERBLINK stars were thus deliberately selected for \emph{K2} monitoring as part of programs aimed at monitoring nearby M dwarfs \citep{Cross16}, as well as stars with large transverse motions, i.e. Galactic halo candidates. Additional SUPERBLINK stars were selected as targets for \emph{K2} after having been proposed by other teams based on different selection criteria whether or not they were known to be high proper motion stars by those teams. We have cross-matched the SUPERBLINK catalog with the final target lists for \emph{K2} campaigns 0-8. This was done by checking the coordinates in the SUPERBLINK catalog against the coordinates provided in the \emph{K2} target lists available online ({\tt http://keplerscience.arc.nasa.gov/K2/}). We used a search radius of 5 arcsec to find matches which identified 27,382 high proper motion SUPERBLINK stars monitored by \emph{Kepler} in the initial \emph{K2} campaigns. For reference, the complete list of SUPERBLINK stars monitored by Kepler is provided in Appendix A. \\ \indent
Figure \ref{fig:position} plots the positions on the sky of these 27,382 SUPERBLINK-K2 (hereafter SBK2) targets from \emph{K2} campaigns 0-8. The black points represent all SBK2 targets, the red circles indicate the fast rotators identified in the present study (see Section \ref{sec:selection}, below). The figure shows that there are more SUPERBLINK stars in campaign 1, 5, 6, and 8 compared to 0, 2, 3, 4, and 7. There are a few reasons for the varying levels of targets per field. Fewer targets were observed in C0 as the mission was still being tested; SUPERBLINK stars were not selected in half of the C4 field for reasons that remain unclear; and fields C2 and C7 contained a fewer number of targets overall as these fields lie near the Galactic plane. Ultimately, SUPERBLINK stars were monitored in each campaign due in large part to the recent increased interest in M-dwarfs among the exoplanet community. Table 1 lists the total number of \emph{K2} targets and the number of long cadence observed SBK2 targets per campaign.\indent

\setlength{\tabcolsep}{0.3in}
\begin{deluxetable}{ccc}
\tablenum{1}
\tablecaption{Number of Targets in \emph{K2} Campaigns 0-8}
\tablehead{
\colhead{Field} & \colhead{\emph{K2}} & \colhead{SBK2} }
\startdata
0 & 7757 & 580 \\ 
1 & 21647 & 6011 \\ 
2 & 13351 & 1395 \\ 
3 & 16348 & 1818 \\ 
4 & 8634 & 2145 \\
5 & 25137 & 3749 \\ 
6 & 28288 & 5407 \\
7 & 13260 & 1809 \\ 
8 & 23564 & 4468
\enddata
\label{tbl:targets}
\end{deluxetable}

We used light curves generated by the {\tt k2phot} photometric pipeline, which is publicly available on GitHub%
\footnote{https://github.com/petigura/k2phot (commit a0d507)}
The light curves are also available on the ExoFOP.
\footnote{https://exofop.ipac.caltech.edu/
}
The reduction corrects for the periodic drift of star images on the \emph{K2} CCD due to roll around the telescope boresight. This data reduction was specifically designed to work with the TERRA exoplanet transit algorithm \citep{Pet13}. \\ \indent
We made two modifications to the {\tt k2phot} light curves in order to eliminate long-term variations in the signals (including instrumental effects) and focus on the detection of short-period modulations expected from stars with short rotation periods. First, we detrended the time series by fitting a third order polynomial to the data and subtracting off the fit. Next, we calculated the mean value of the flux over the entire \emph{K2} light curve, then subtracted that value to obtain a time series of the residuals with arbitrary flux units, $f$. Next, we rejected all points that deviated $f$ $\geq$ $3\%$ or $f$ $\leq$ $-30\%$ from the mean in order to exclude instrumental artifacts and intrinsic bursts such as flares. We chose a deep threshold for rejection as we did not want to remove any potential eclipses that could be investigated in future work. We kept an account of all points that were rejected and flagged these points as possible flare events (outliers above the mean) or eclipses (outliers below the mean). Many of these rejections appear to be spurious spikes (possibly instrumental artifacts in many cases) with most light curves showing $\sim$10 of these spurious events. We then ran these ``cleaned'' light curves through our cross-correlation algorithm (which does not require uniform sampling - see below) to identify periodic signals.

\section{Identification of Periodic and Quasi-Periodic Modulations}
\label{sec:acf}
Lomb-Scargle periodigrams are widely used to analyze time series data to find rotation periods of stars \citep{Scarg}. The strength of Lomb-Scargle analysis is that it does not require the signal to be evenly sampled. However, the signal to be detected must be very nearly periodic over the entire time series. Before \emph{Kepler}, this was the preferred method to search for rotation signals in stellar light curves because most data sets were irregularly sampled over the course of a few nights of observations \citep{McQ13}. A potential drawback of the method is that rotation modulations is most stars are not strictly periodic because stars spot patterns change and flares or transit events occur. \\ \indent
Now, with publicly available \emph{K2} data, we can explore more options for time series analysis. \citet{McQ13} thoroughly demonstrated that the auto-correlation function (ACF) is more powerful when used with \emph{Kepler} data because it is a measure of the self-similarity of the light curve over characteristic timescales and does not require the signal to be strictly repeating. In this paper, we adopt a similar procedure to identify stellar fast rotators in the \emph{K2} data and estimate their rotation periods. \\ \indent
 Assuming a lag $k$ in the evenly sampled time-series, we define the auto-correlation coefficient, $r_k$, to be

\begin{equation}
r_k = \frac{\sum_{i=1}^{N-k}(f_i-\bar{f})(f_{i+k}-\bar{f})}{\sum_{i=1}^{N-k}(f_i-\bar{f})^2},
\end{equation} 

where $\bar{f}$ is the mean flux value, $f_i$ is the flux value at the $i$th point in the time series, $f_{i+k}$ is the flux value at the $i$+$k$th value in the time series, and N is the total number of sample points. The ACF itself is all coefficients, $r_k$, plotted against their respective $k$ values multiplied by the cadence, $c$, which provides a time delay rather than a pixel lag. In our analysis, we search for modulations with periods that have up to half the length of the time series. The ``rotation period'' of a star is determined from the characteristic timescale of the variations in the \emph{K2} light curve from:

\begin{equation}
P_{rot} = k_{max}*c
\end{equation} 

where $k_{max}$ is the location of the first peak in the auto-correlation function, $r_k$. In general, a strictly periodic and infinitely repeating signal will also display significant peaks at lags of $2k_{max}$, $3k_{max}$, ..., $nk_{max}$ all being aliases of the fundamental period. A recurrent signal that changes its fundamental pattern over time, on the other hand, will have alias peaks that are weaker than the first, and they may not peak at exact integer values of the fundamental period. Determining the location of the first peak proves to be difficult using a simple search for the first local maximum. That is because the ACF typically also displays high frequency, low amplitude signals that produce many hundreds of local maxima. For \emph{K2} light curves, a simple box-smoothing of the ACF is generally found to be insufficient in removing the high frequency signal as the size of the box would have to differ for each target so that the intrinsic signal would not be degraded by the smoothing. Optimally, the size of the box would need to correspond to the fundamental frequency of the ACF, i.e. the rotation period of the star, a value that is not known a priori. \\ \indent
As an alternative method to identify the astronomically relevant signals, we have adopted the use of the Fast Fourier Transform (FFT) of the ACF. The advantage is that the fundamental frequency of the signal with at least a few of its aliases in the ACF will present itself as as a relatively strong signal in the FFT. We thus compute the FFT of all the ACF from the \emph{K2} light curves, and search for maxima in those FFT.  \\ \indent
Since the signal we are after may not be strictly repeating, the aliases in the ACF may also not be evenly spaced, and the period extracted from the FFT generally only yields an {\em approximate} estimate of the first peak in the ACF. In addition, we find that the FFT has limited accuracy in measuring the periods from low-frequency signals. We therefore use the period identified in the FFT as an {\em initial guess} to identify the fundamental first peak in the ACF. We then perform a search for local maxima in the ACF using a parabola fit around the initial guess. We perform a fit of a section of the ACF that is centered on the location of the initial guess for the first peak with a width of one half of the period corresponding to that fundamental first peak. We use the three coefficients $a_0$, $b_0$, and $c_0$ returned from the polynomial fit ($a_0 x^2$+$b_0 x$+$c_0$ ) to calculate the center, i.e. the effective stellar rotation period $P_{rot,0}$ associated with the modulation; the peak, i.e. amplitude of the ACF, $A_0$; and the parabola opening, $B_0$. These three values are calculated via the following equations..

\begin{equation}
P_{rot,0} = \frac{-b_0}{2a_0} 
\end{equation} \\ 
\begin{equation}
A_0 = c_0 - \frac{b_0^2}{4a_0}
\end{equation} \\ 
\begin{equation}
B_0 = a_0P_{rot,0}^2
\end{equation} 

In addition, we also locate the first four aliases in the ACF based on the initial guess, and perform similar second-order polynomial fits to estimate the amplitudes and parabola openings of those peaks, and their associated effective stellar rotation periods. Thus for each alias $k$, we find the coefficient $a_k$, $b_k$, and $c_k$, and compute:

\begin{equation}
P_{rot,k} = \frac{-b_k}{2a_k} 
\end{equation} \\ 
\begin{equation}
A_k = c_k - \frac{b_k^2}{4a_k}
\end{equation} \\ 
\begin{equation}
B_k = a_kP_{rot,k}^2
\end{equation} 

Finally, we calculate the average values of these parameters from the fundamental peak and its first four aliases:

\begin{equation}
P_{rot} = \frac{1}{5}\sum^5_{k=1} P_{rot,k}
\end{equation} \\
\begin{equation}
A = \frac{1}{5}\sum^5_{k=1} A_k
\end{equation} \\
\begin{equation}
B = \frac{1}{5}\sum^5_{k=1} B_k
\end{equation}

\begin{figure}[t]
\centering
  \includegraphics[width=\linewidth,height=170mm]{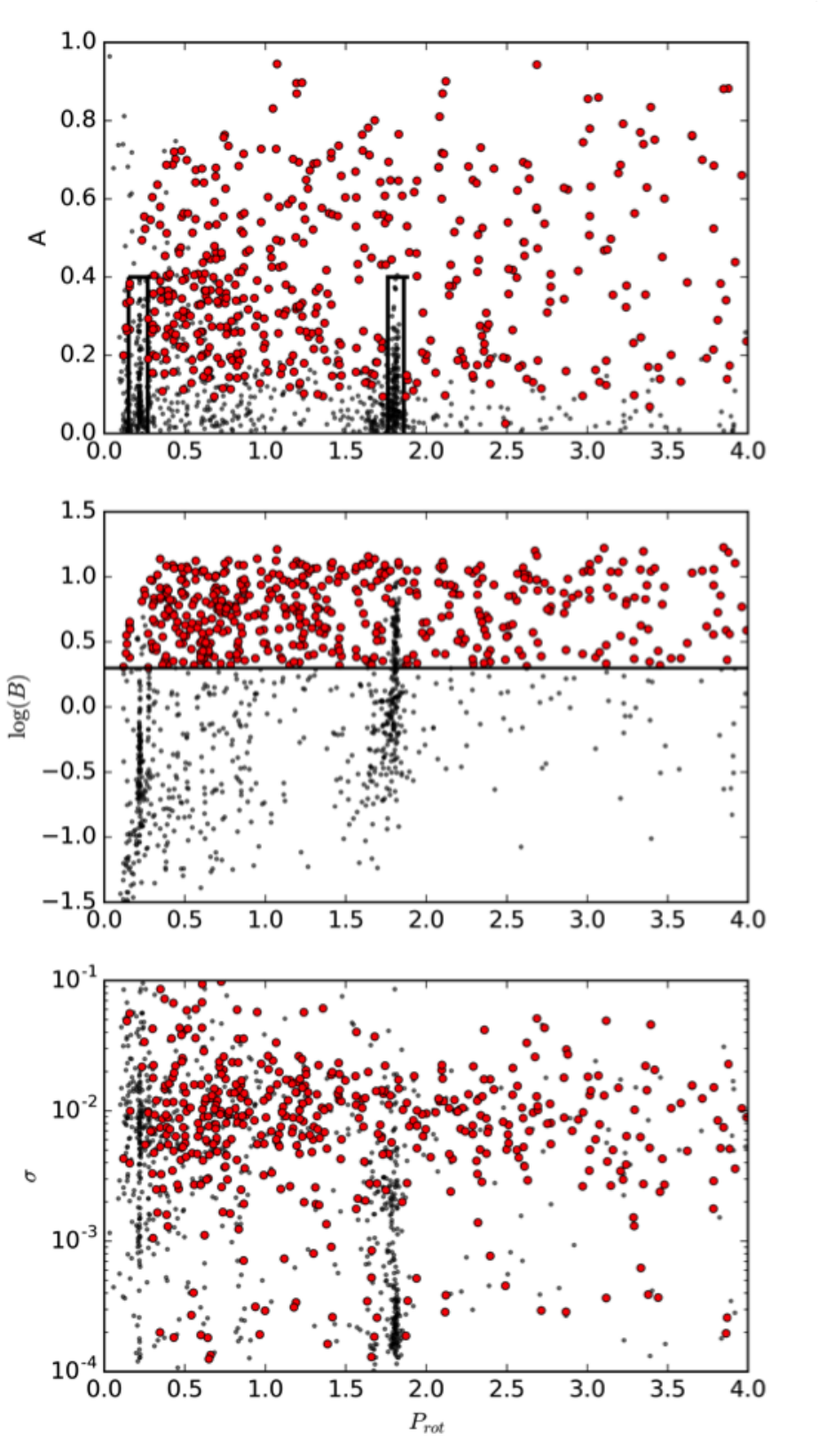}
\caption{The parameters A and B calculated from the ACF used to reject false positive rotation signals. The top panel indicates the box rejection of targets with $P_{rot}$ $\sim$ $0.22$ days and $P_{rot}$ $\sim$ $1.75$ days with $A$ $\leq$ $0.4$. The middle panel indicates the rejection of any star with $B$ $<$ $2$. These selection criteria were determined after careful ``by-eye'' examination of hundreds of \emph{K2} light curves. The bottom panel plots the amplitude of the light curve against the stellar rotation period. There does not appear to be correlation between amplitude and rotation period.}
\label{fig:cuts}
\end{figure}

\begin{figure*}
\subfigure{
  \includegraphics[scale=0.3]{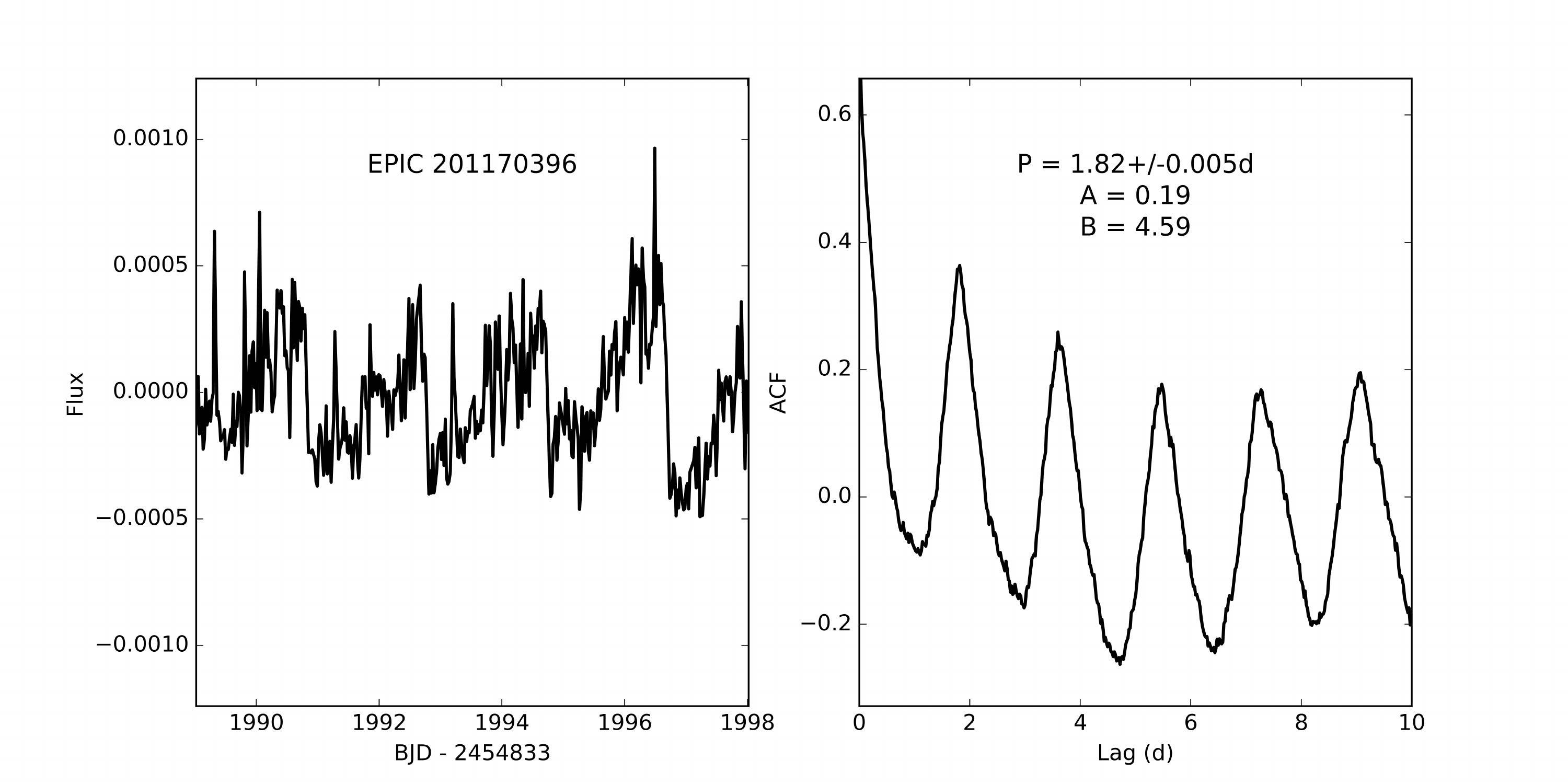}
}
\subfigure{
  \includegraphics[scale=0.3]{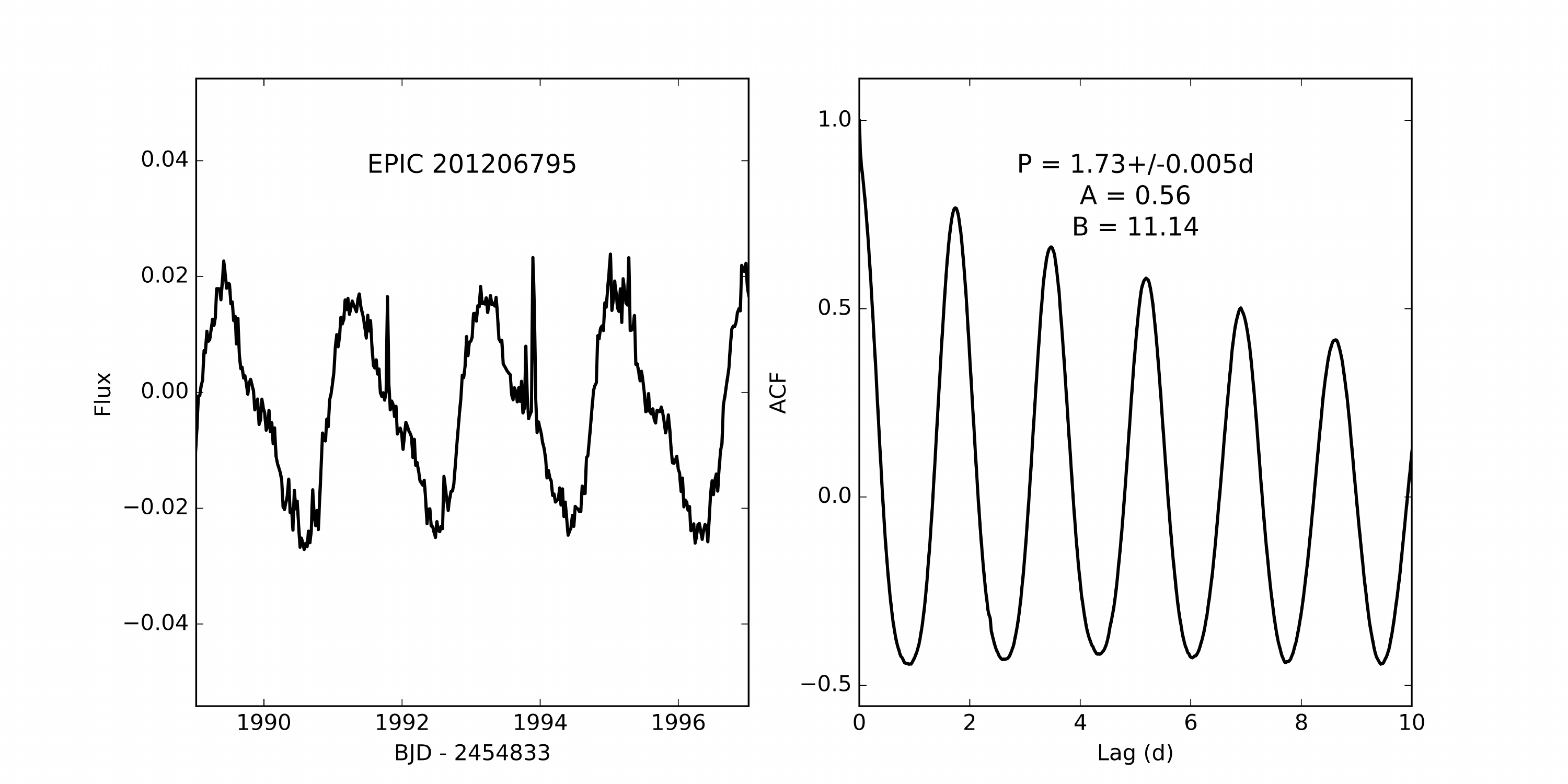}
  }
\caption{Two examples of targets with $P_{rot}$ $\sim$ $1.75$ days, at which we identify a contaminating instrumental signal in a number of stars. The left two panels shows a star with the contaminating signal, which follows a characteristic ``sawtooth'' pattern. The star on the right, though its estimated period is close to the contaminating signal, shows a pattern more consistent with stellar rotation. We distinguish the two from the amplitude of their ACF: stars with lower correlation amplitude ($A$ $<$ $0.4$) are rejected as contaminated, stars with higher correlation amplitudes ($A$ $>$ $0.4$) are saved.}
\label{fig:example_rightbox}
\end{figure*}

\begin{figure*}
\subfigure{
  \includegraphics[scale=0.3]{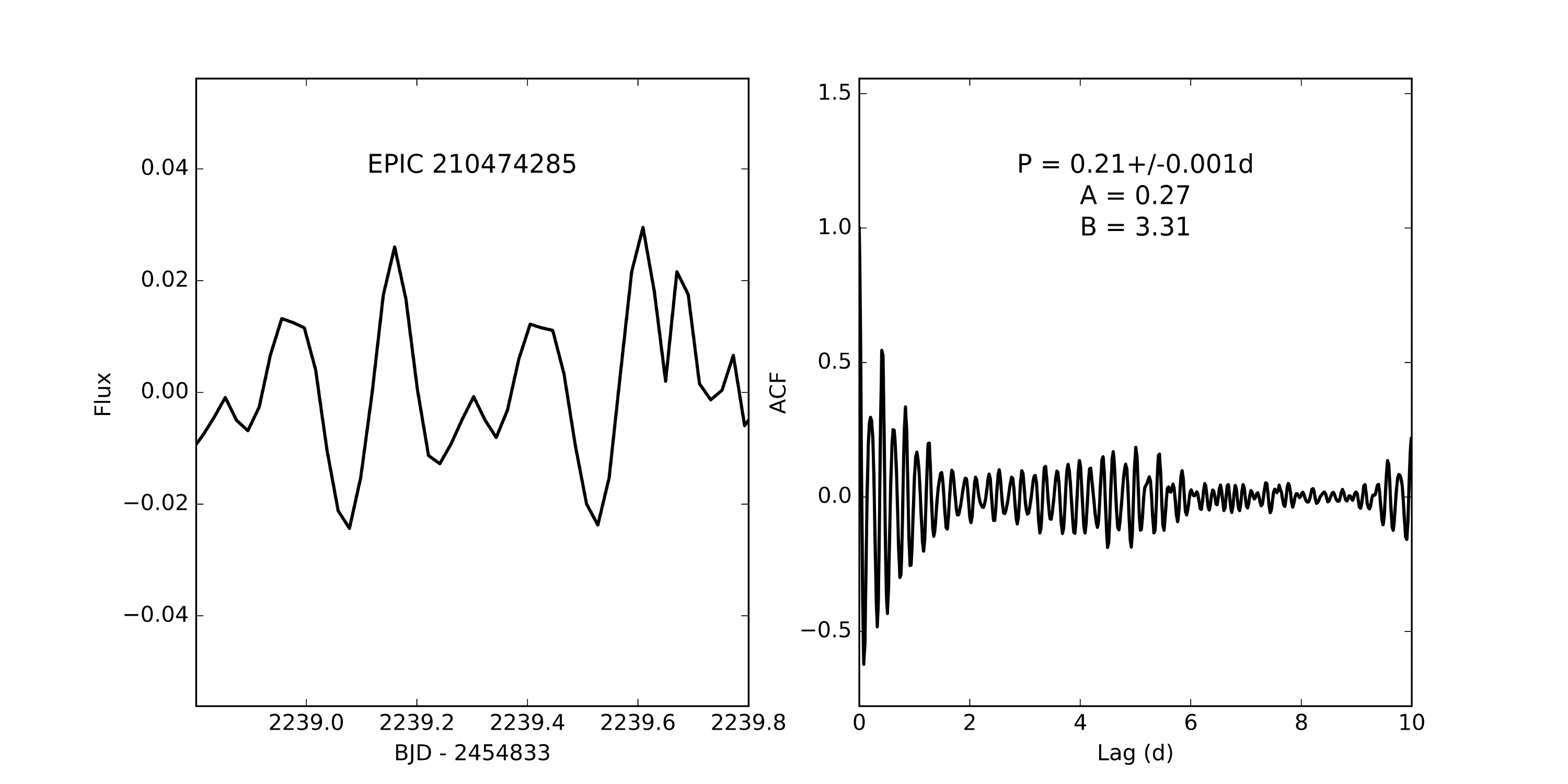}
  }
\subfigure{
  \includegraphics[scale=0.3]{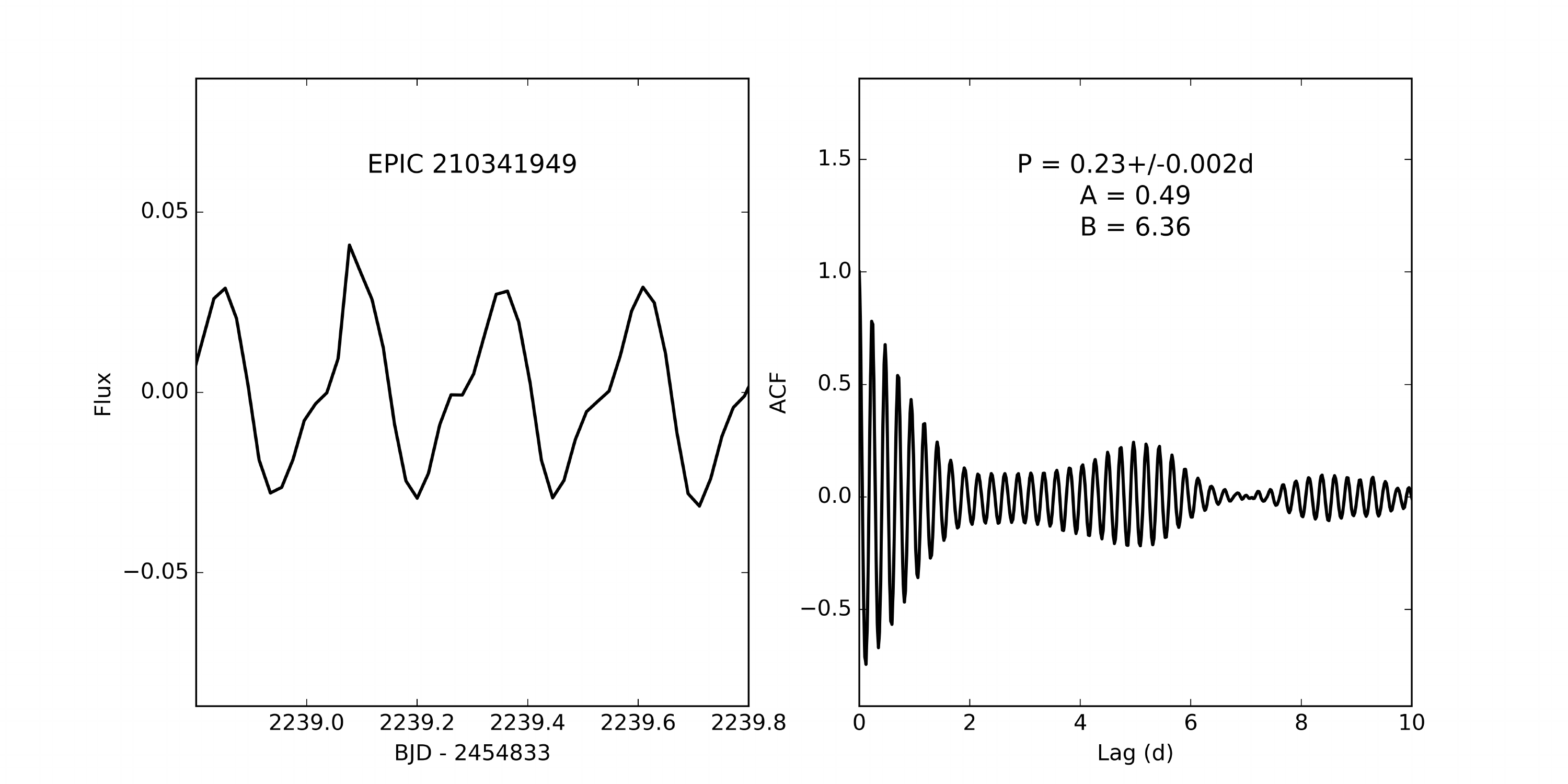}
  }
\caption{Two examples of targets where our analysis suggests recurrent signals with periods $\sim$ $0.23$ days. The left two panel shows a target with a characteristic ``jagged'' pattern in the light curve, yielding a value of $A$ $<$ $0.4$. We assume that the signal in these stars is likely an instrumental artifact, and all similar objects are rejected from the analysis. The star in the right two panels also has a recurrent signal with a period $\sim$ $0.23$ days, however the strength of the correlation peaks $A$ $>$ $0.4$ suggests the signal is likely intrinsic to the star, and these objects are kept in database.}
\label{fig:example_leftbox}
\end{figure*}

Values of the A and B parameters calculated for all the stars with detected modulations are shown in Figure \ref{fig:cuts}. A first look of these parameters revealed the appearance of a systematic false positive at $P_{rot}\sim0.22$ days and $P_{rot}\sim1.75$ days. This systematic false positive appears in all campaigns although some are affected more than others. After trial and error, we determined the best way to ensure we do not select these false positives for the final sample was to reject all targets with measured periods within $\pm 0.08$days of the values above, and with $A\leq0.4$. This rejection criterion is shown in the top panel of Figure \ref{fig:cuts}. The rejected targets are in black points while the targets that pass this selection criterion are plotted in red points. Examples of targets that pass and fail this rejection are shown in Figures \ref{fig:example_rightbox} and \ref{fig:example_leftbox}. \indent
After we make these box rejections, we then impose the following selection criteria. 

\begin{equation}
0.0 < P_{rot} \leq 4.0  
\end{equation} \\ 
\begin{equation}
A > 0.0 \hspace{0.5cm}  \forall k=0,5
\end{equation} \\ 
\begin{equation}
B > 2.0
\end{equation}

\begin{figure*}
\subfigure{
  \includegraphics[scale=0.3]{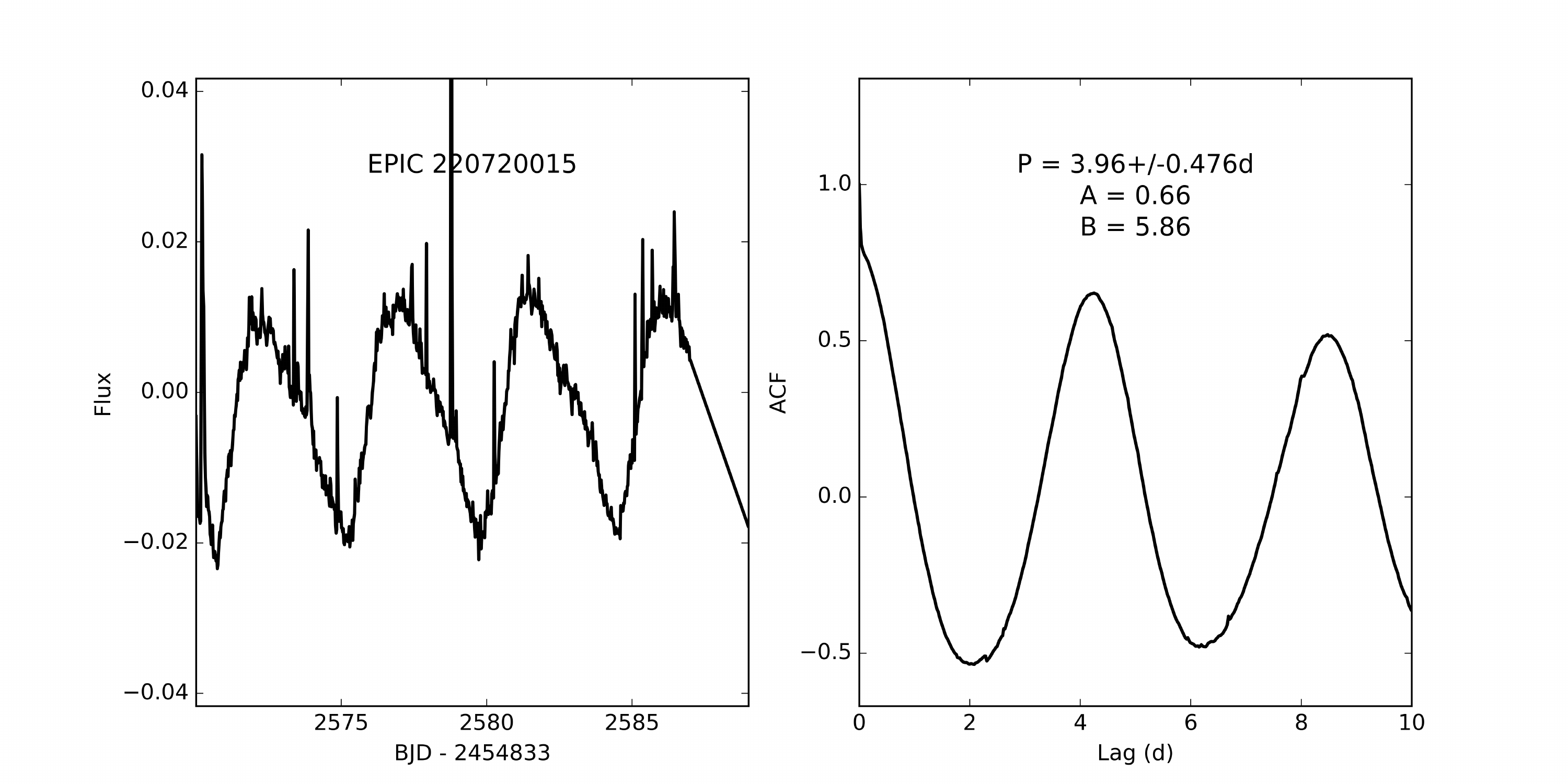}
    }
\subfigure{
  \includegraphics[scale=0.3]{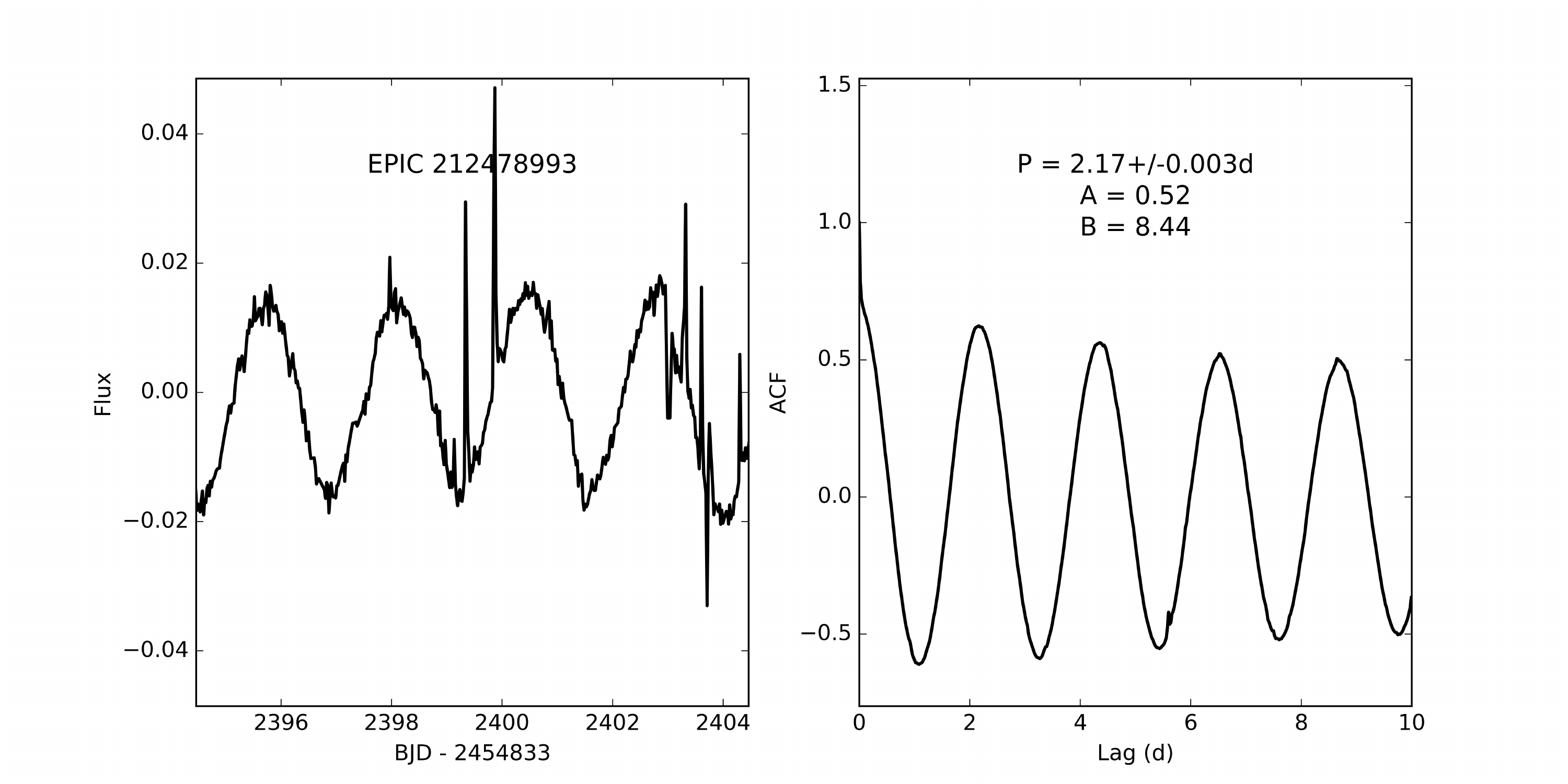}
}
\caption{Two examples of light curves with a strong correlation signal. The high B values pass our selection criteria and is above the threshold to be considered a high confidence result.}
\label{fig:example_strong}
\end{figure*}

\begin{figure*}
\subfigure{
  \includegraphics[scale=0.3]{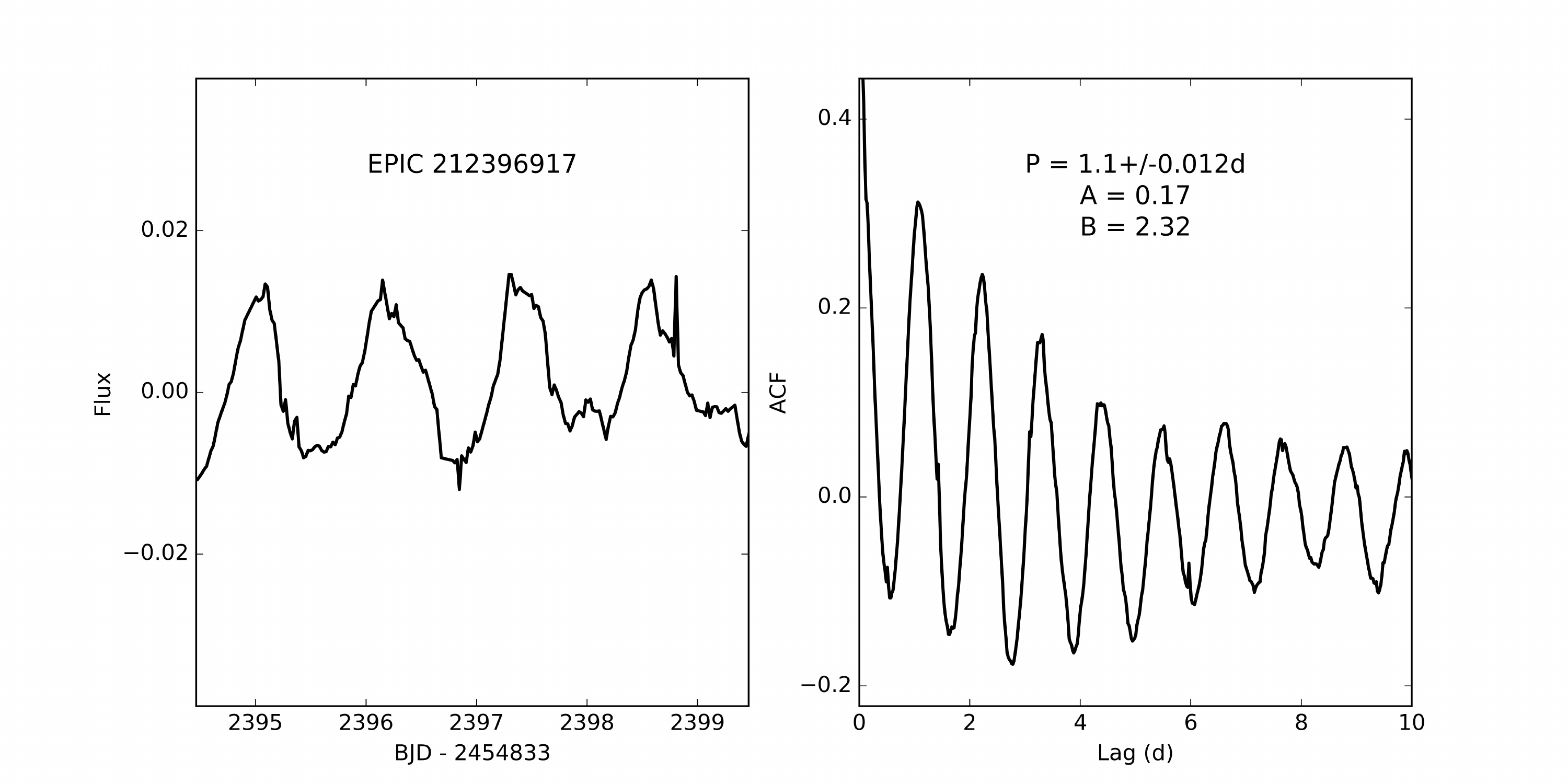}
}
\subfigure{
  \includegraphics[scale=0.3]{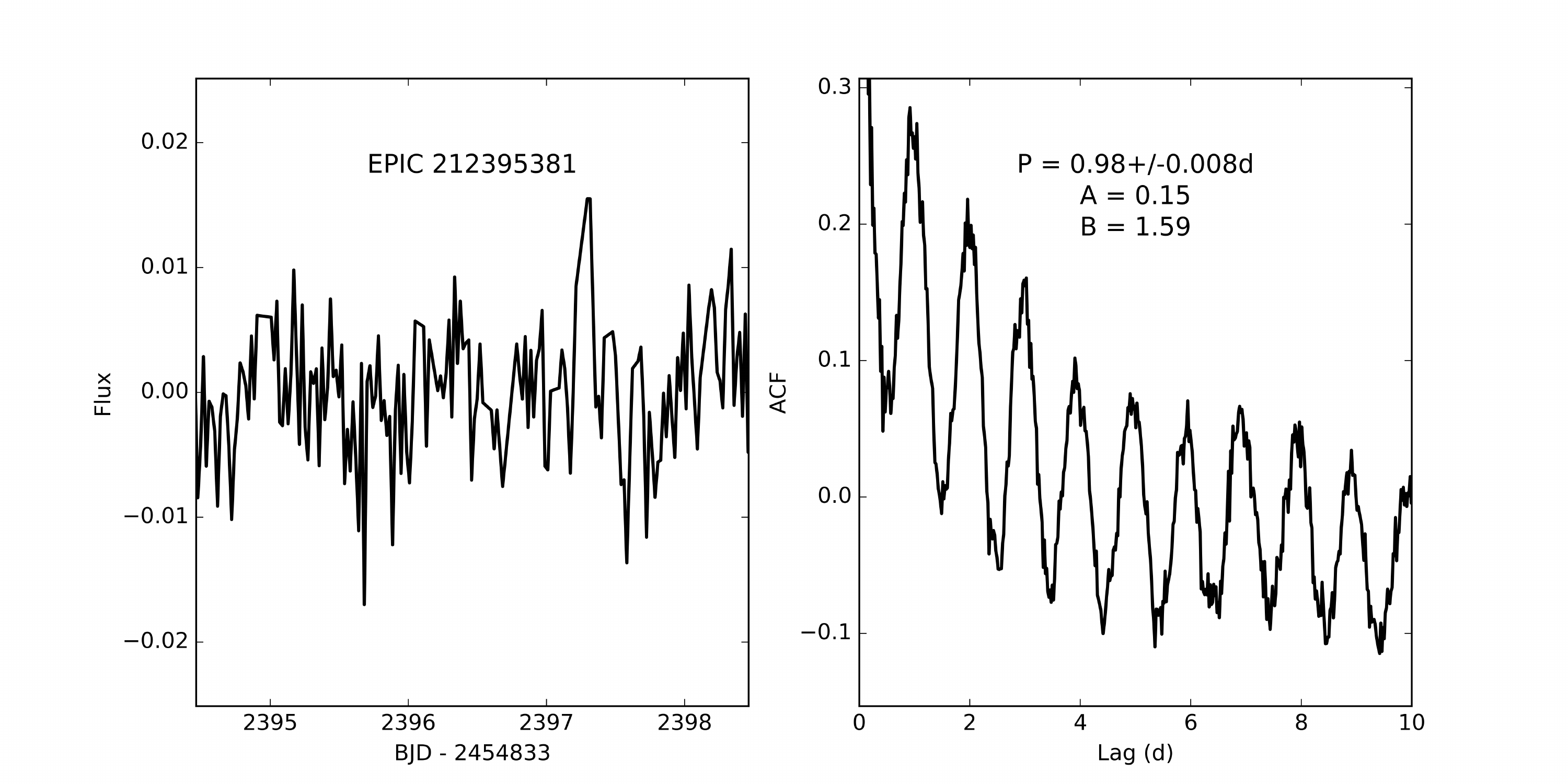}
}
\caption{Two example of light curves with B values close to the limit of our selection cut off ($B$ $>$ $2.0$). EPIC 212396917 (top) has a clear recurrent signal in its light curve, while EPIC 212395381 (bottom) is relatively noisier with no clearly identifiable pattern. While their auto-correlation functions right panels) appear similar at first glance, EPIC 212395381 has peaks of lower amplitude, which places its values of A and B outside of our selection limit.}
\label{fig:example_medium}
\end{figure*}

The first criterion simply defines our pre-determined goal of identifying modulations from fast rotators. The second criterion requires that the fundamental peak and its first four aliases all have a positive amplitude, which essentially requires that any signal should have a minimal number of well-defined aliases, as one would expect from a modulation that persists for at least a few cycles before changing significantly. The third selection criterion is a requirement that the fundamental peak and its first four aliases be the dominant signals in the ACF, i.e. have widths that are comparable with the spacing between them. The middle panel of Figure \ref{fig:cuts} shows the cutoff in $\log B$ vs $P_{rot}$ space. The specific values of B that are associated with ``clean'' signals in the ACF are subjective, but is based on the visual inspection of several hundreds of ACF from \emph{K2} light curves. A few typical examples are shown in Figures \ref{fig:example_strong} and \ref{fig:example_medium}. \indent

\begin{figure*}
\subfigure{
  \includegraphics[scale=0.3]{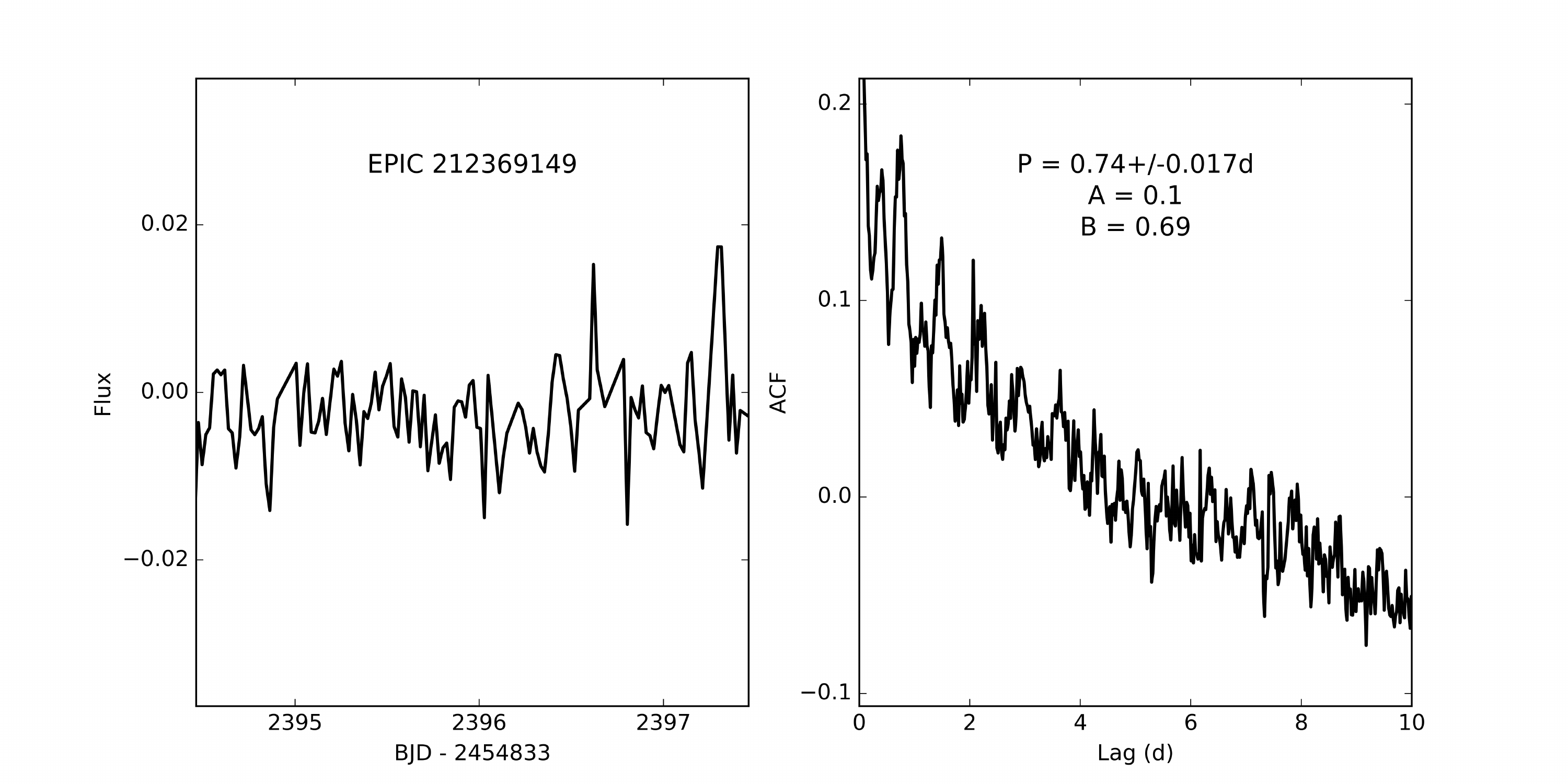}
}
\subfigure{
  \includegraphics[scale=0.3]{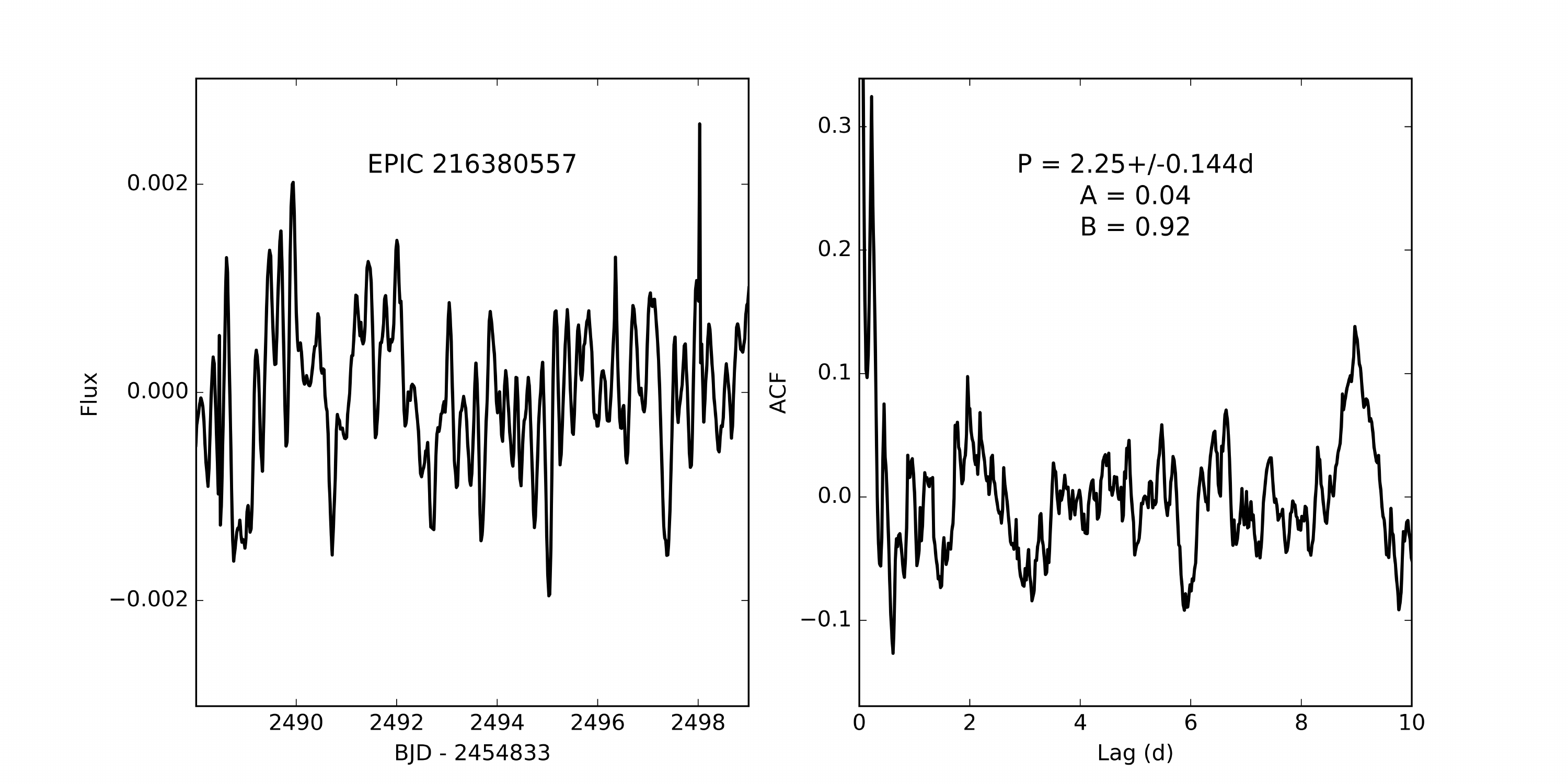}
}
\caption{Two examples of noisy lights curves, where no clear modulation can be detected despite the identification of a signal in the auto-correlation function. Although the auto-correlation functions (panels on the right) show a number of local maxima, those are not well defined which is reflected in the low values of A and B obtained from a parabola fit of the first five local maxima. We thus reject all stars that have similarly low values of B, setting the limit for positive detection at $B$ $>$ $2$.}
\label{fig:example_weak}
\end{figure*}

In general, we find that ACF with recovered values of $B$ $>$ $2$ correspond to light curves in which a periodic modulation is noticeable on a ``by-eye'' inspection of the light curve, and for which we have a good level of confidence. ACF with $0$ $<$ $B$ $<$ $2$, on the other hand, correspond to extremely noisy light curves that appear to be dominated with stochastic signals, and for which an intrinsic modulation is not readily apparent upon visual inspection, see Figure \ref{fig:example_weak}. Finally, ACF with recovered values of $1$ $<$ $B$ $<$ $2$ correspond to light curves in which a by-eye examination suggests a possible weak modulation, but we cannot be certain about it. We therefore place a higher confidence on those targets with $B>2$, and only allow these targets in to the final sample; these stars are the ones plotted as candidate fast rotations in Figures \ref{fig:position} and \ref{fig:cuts} (red symbols). Our method of peak selection differs from earlier studies, but it yields results that are comparable to previous analyses \citep{McQ14,Arm16}. 

\section{Identification of Fast Rotators}
\label{sec:selection}
Young ($<$ $150$ Myr), early type M dwarfs typically have rotation periods $P_{rot}$ $<$ $4$ days \citep{MH08}. We have thus focused our analysis on the identification of SBK2 stars with these rotation rates. \\ \indent
Fast rotators are identified from our combined ACF and FFT analysis as described above, with the FFT identifying regular peaks in the ACF corresponding to a recurrent signal and its aliases, and with individual fits to  these peaks used to compute the associated stellar rotation period and assess the quality of the identification. This combined ACF and FFT algorithm identified 508 candidate fast rotators in the SUPERBLINK stars monitored in \emph{K2} campaigns 0-8. \indent

\setlength{\tabcolsep}{0.04in}
\begin{deluxetable*}{lrrrrrrrrrrrrrr}
\tablenum{2}
\tabletypesize{\footnotesize}
\tablecaption{Low-mass Fast Rotators in \emph{K2} Campaigns 0-8}
\tablewidth{0pt}
\tablehead{
\colhead{SB} & \colhead{EPIC} & \colhead{$\alpha$} & \colhead{$\delta$}  & \colhead{$pm_{\alpha}$}   & \colhead{$pm_{\delta}$}  & \colhead{NUV} & \colhead{V}  & \colhead{V-J}  &  \colhead{$P_{rot}$}  & \colhead{$\sigma$ $P_{rot}$} & \colhead{A \tablenotemark{1}} & \colhead{B \tablenotemark{2}} &  \colhead{$\sigma$ \tablenotemark{3}} &  \colhead{Field}   \\
& & \colhead{(J2000)} & \colhead{(J2000)} & \colhead{(mas/yr)} & \colhead{(mas/yr)} & \colhead{(mag)} & \colhead{(mag) }& \colhead{(mag)} & \colhead{(days)} & \colhead{(days)}  }
\startdata
PM I00389$+$0828	&	220574720	&	9.746918	&	8.478116	&	0.1000	&	-0.0471	&	18.97	&	10.02	&	2.11	&	1.25	&	0.002	&	0.65	&	12.1	&	0.019	&	8	\\
PM I00421$+$0509	&	220422705	&	10.525499	&	5.156477	&	0.0152	&	-0.0103	&	12.34	&	12.38	&	-0.15	&	3.78	&	0.095	&	0.22	&	3.6	&	0.002	&	8	\\
PM I00440$+$0932	&	220623236	&	11.005507	&	9.549399	&	0.0232	&	-0.0468	&	13.16	&	10.18	&	1.73	&	0.39	&	0.001	&	0.69	&	12.2	&	0.012	&	8	\\
PM I00452$+$0423	&	220383711	&	11.307917	&	4.388321	&	0.0428	&	-0.0199	&	99.99	&	18.89	&	4.92	&	3.99	&	0.012	&	0.24	&	3.9	&	0.009	&	8	\\
PM I00460$+$0757	&	220552625	&	11.524994	&	7.957099	&	0.1190	&	0.0204	&	20.97	&	17.80	&	5.00	&	1.07	&	0.000	&	0.50	&	8.2	&	0.006	&	8	\\
PM I00499$+$0708	&	220516443	&	12.488429	&	7.143973	&	0.1830	&	0.0066	&	99.99	&	16.07	&	4.66	&	0.65	&	0.001	&	0.53	&	10.1	&	0.009	&	8	\\
PM I00502$+$0837	&	220581451	&	12.573019	&	8.626148	&	0.0596	&	-0.0303	&	99.99	&	14.70	&	4.96	&	1.32	&	0.005	&	0.10	&	5.6	&	0.007	&	8	\\
PM I00523$+$0638	&	220494537	&	13.078835	&	6.640716	&	-0.0459	&	-0.0360	&	99.99	&	17.85	&	4.60	&	0.51	&	0.003	&	0.28	&	3.5	&	0.004	&	8	\\
PM I00526$+$0337W	&	220346833	&	13.155402	&	3.631029	&	-0.0247	&	-0.0469	&	99.99	&	18.19	&	4.70	&	1.37	&	0.001	&	0.28	&	4.1	&	0.007	&	8	\\
PM I00532$+$1056	&	220680367	&	13.305914	&	10.938226	&	0.0379	&	-0.0365	&	99.99	&	15.03	&	3.12	&	0.74	&	0.004	&	0.76	&	10.7	&	0.026	&	8	\\
PM I00532$+$0017	&	220208154	&	13.316236	&	0.289406	&	0.0513	&	-0.0749	&	23.71	&	15.52	&	3.13	&	1.64	&	0.000	&	0.78	&	14.4	&	0.020	&	8	\\
PM I00535$+$0725	&	220529150	&	13.396629	&	7.417181	&	0.0590	&	-0.0317	&	99.99	&	17.09	&	4.40	&	0.81	&	0.000	&	0.26	&	4.5	&	0.014	&	8	\\
PM I00551$+$0011	&	220205547	&	13.796958	&	0.193448	&	0.0393	&	-0.0631	&	99.99	&	19.20	&	4.90	&	0.85	&	0.001	&	0.41	&	6.6	&	0.017	&	8	\\
PM I00552$+$0531	&	220441862	&	13.822223	&	5.524729	&	0.0476	&	-0.0246	&	99.99	&	16.53	&	3.70	&	2.57	&	0.003	&	0.62	&	10.6	&	0.016	&	8	\\
PM I00553$+$0235	&	220298113	&	13.829959	&	2.589331	&	-0.0866	&	-0.1156	&	23.52	&	16.03	&	3.88	&	2.94	&	0.008	&	0.42	&	7.4	&	0.009	&	8	\\
PM I00560$+$0249W	&	220308656	&	14.003282	&	2.822468	&	-0.0738	&	-0.0792	&	21.69	&	13.44	&	3.60	&	0.67	&	0.024	&	0.27	&	4.8	&	0.013	&	8	\\
PM I00570$+$0223	&	220288486	&	14.252249	&	2.391407	&	-0.0920	&	-0.0909	&	21.75	&	15.30	&	4.52	&	0.58	&	0.003	&	0.26	&	4.0	&	0.005	&	8	\\
PM I00577$+$0805	&	220558588	&	14.434575	&	8.097114	&	-0.0127	&	0.0659	&	99.99	&	16.69	&	5.10	&	0.32	&	0.001	&	0.20	&	2.9	&	0.002	&	8	\\
PM I00578$+$0440	&	220398184	&	14.472344	&	4.680145	&	0.0857	&	0.0042	&	22.57	&	14.63	&	3.31	&	1.24	&	0.003	&	0.10	&	3.7	&	0.057	&	8	\\
PM I00583$+$1203	&	220720015	&	14.586460	&	12.056598	&	-0.0332	&	-0.0306	&	99.99	&	16.69	&	4.53	&	3.96	&	0.476	&	0.66	&	5.9	&	0.010	&	8	
\enddata
\tablenotetext{1}{Average correlation value for the first five maxima in the auto-correlation function.}
\tablenotetext{2}{Parameter indicating the dominance of the auto-correlation peak (see text).}
\tablenotetext{3}{Standard deviation in the light curve after detrending, showing the amplitude of the variability signal. }
\tablecomments{The full table is available in the electronic version of the paper.}
\label{tbl:fastrot}
\end{deluxetable*}

\begin{figure}[]
\centering
  \includegraphics[width=\linewidth,height=75mm]{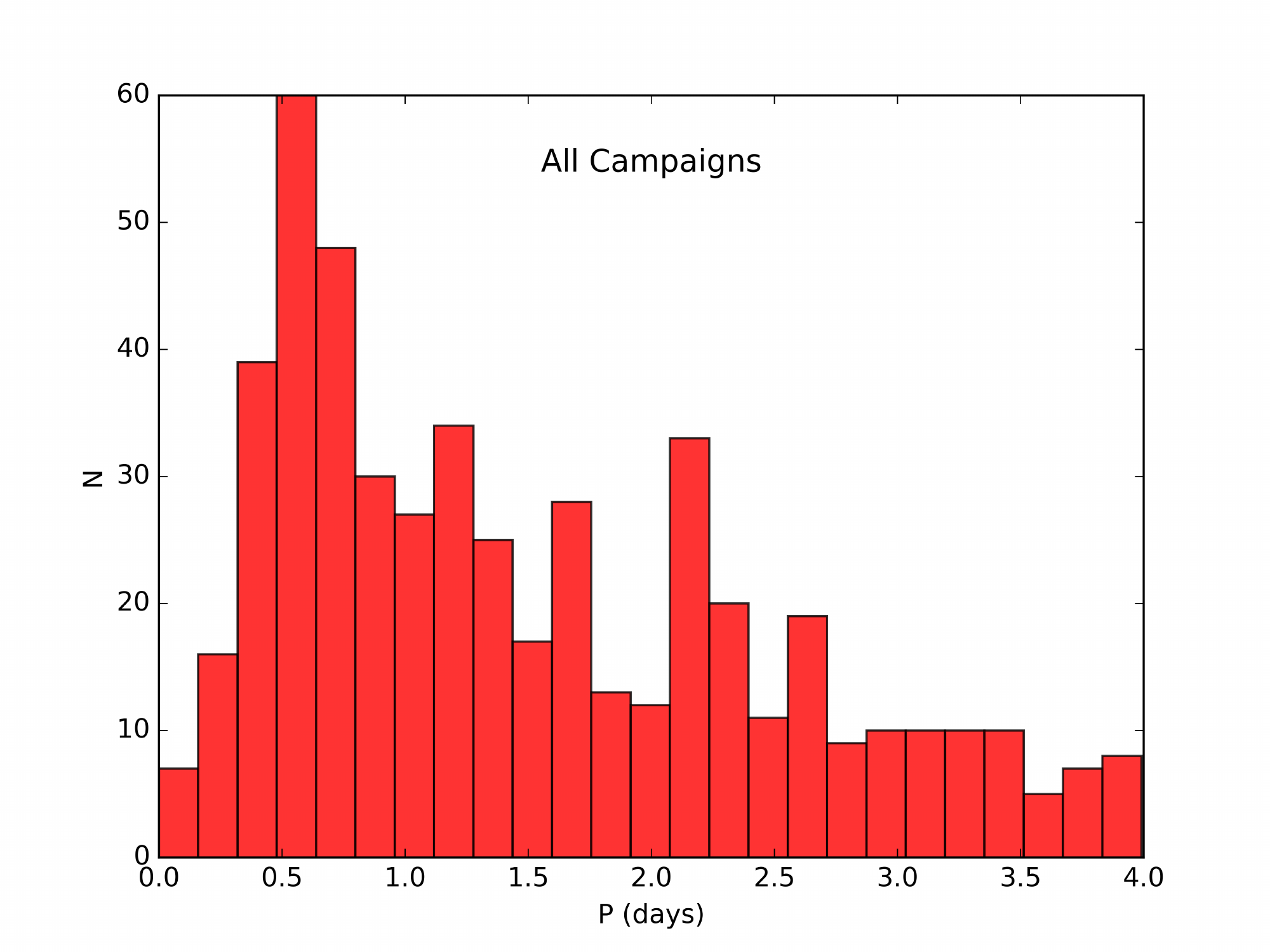}
\caption{Distribution of the rotation periods identified in this work. We find fewer targets with with $P_{rot}$ $>$ $1.5$ days. This may be intrinsic to the population of young stars we study here or a systematic trend in the data. In future work we would like to investigate these possibilities and improve our pipeline to find rotation rates up to 4 days.}
\label{fig:histogram}
\end{figure}

\begin{figure}
\centering
  \includegraphics[width=\linewidth,height=75mm]{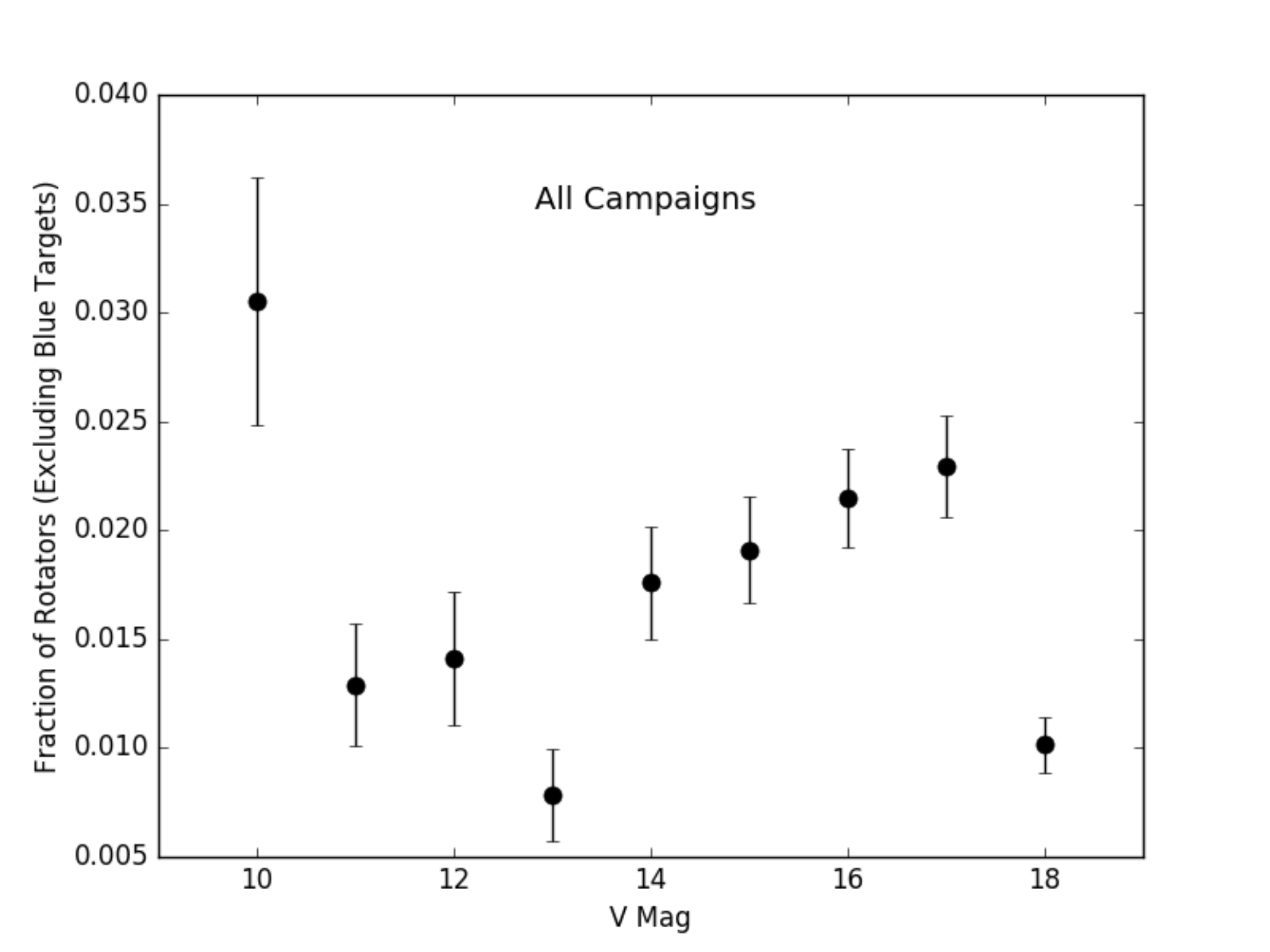}
\caption{Fraction of rapidly rotating M dwarfs as a function of magnitude. At fainter magnitudes, the fraction of fast rotators increases because the fainter magnitudes are comprised mainly of M dwarfs as indicated by Figure \ref{fig:rpm}. The drop off at $V$ $=$ $18$ is indicative of the \emph{K2} detection limit for fast rotators. This limit is due to light curves that are too noisy to detect a repeating signal.}
\label{fig:fractionofrotators}
\end{figure}

Figure \ref{fig:histogram} shows a histogram of the distribution of candidate fast rotators as a function of their assumed, estimated rotation period, $P_{rot}$. The number of detected fast rotators decreases at the rotation period increases, with most objects having estimated rotation periods $P_{rot}$ $<$ $1.5$ days. There may be several explanations for this. The distribution may reflect the true statistics of fast rotators, as defined by the physical mechanisms responsible for their slow down. On the other hand, slower rotators have smaller spots, and thus their modulations would have lower intrinsic amplitudes, which would make them harder to identify amidst the instrumental noise. If this is true, then the identification of stars with rotation periods $P_{rot}$ $\gtrsim$ $2$ days might therefore be a significant challenge using \emph{K2} data. In future work we would would like to develop an algorithm that can pick out more stars with rotation periods $P_{rot}$ $>$ $2$ days, and especially stars with $P_{rot}$ $>$ $4$ days, for which we did not look here. \\ \indent
Figure \ref{fig:fractionofrotators} shows the fraction of fast rotators identified in each magnitude bin. The trend indicates an increase in detection of fast rotators at fainter magnitudes, which is counterintuitive because one would expect rotation modulations to be harder to detect in fainter stars. However, this occurs because the fainter targets are comprised mainly of M dwarfs, while the brighter targets are largely FGK dwarfs (see Figure \ref{fig:rpm}). The age-activity relationship presented by \citet{West08} indicates that M stars remain active longer than FGK stars. Therefore, late type stars will persist as rapid rotators longer than early type stars making it more likely to find a rapid rotator in a sample dominated by M dwarfs like the one we present here. Finally, the trend in Figure \ref{fig:fractionofrotators} reveals a sharp downturn at $V$ $=$ $18$, which suggests that $V$ $=$ $18$ is the magnitude limit for detecting modulations from spots in \emph{K2} data. This limit is due to the light curves becoming too noisy to detect a repeating signal. \indent

The complete list of SBK2 targets identified as fast rotators can be found in Table 2. The first 20 lines of the table are shown in the print version; the complete list is available in the electronic version of the paper. \\ \indent
Columns 1 and 2 give the identification number of the star in the SUPERBLINK and EPIC catalogs, respectively. Columns 3 and 4 list the J2000 right ascension and declination, respectively. Columns 5 and 6 list the SUPERBLINK catalog proper motion in the direction of right ascension and declination, respectively, both in units of seconds of arc per year. Column 7 lists NUV magnitude from \emph{GALEX}, when available. Column 8 lists V magnitudes from SUPERBLINK. Column 9 lists V-J color using 2MASS infrared magnitudes, J. Columns 10 and 11 list the rotation period and uncertainty in rotation period calculated in this work. Columns 12 and 13 lists the A and B values discussed in Section \ref{sec:acf} that were used to differentiate between false positive and real detections. Column 14 is a measure of the amplitude of the intrinsic variability of the star, $\sigma$, computed by taking the standard deviation of the signal after our detrending procedure. Column 15 is the \emph{K2} field corresponding to the target. 

\section{Kinematic Analysis of the Fast Rotators}
\label{sec:kinematics}
\subsection{Reduced Proper Motions}
\begin{figure}[]
\centering
  \includegraphics[width=\linewidth,height=70mm]{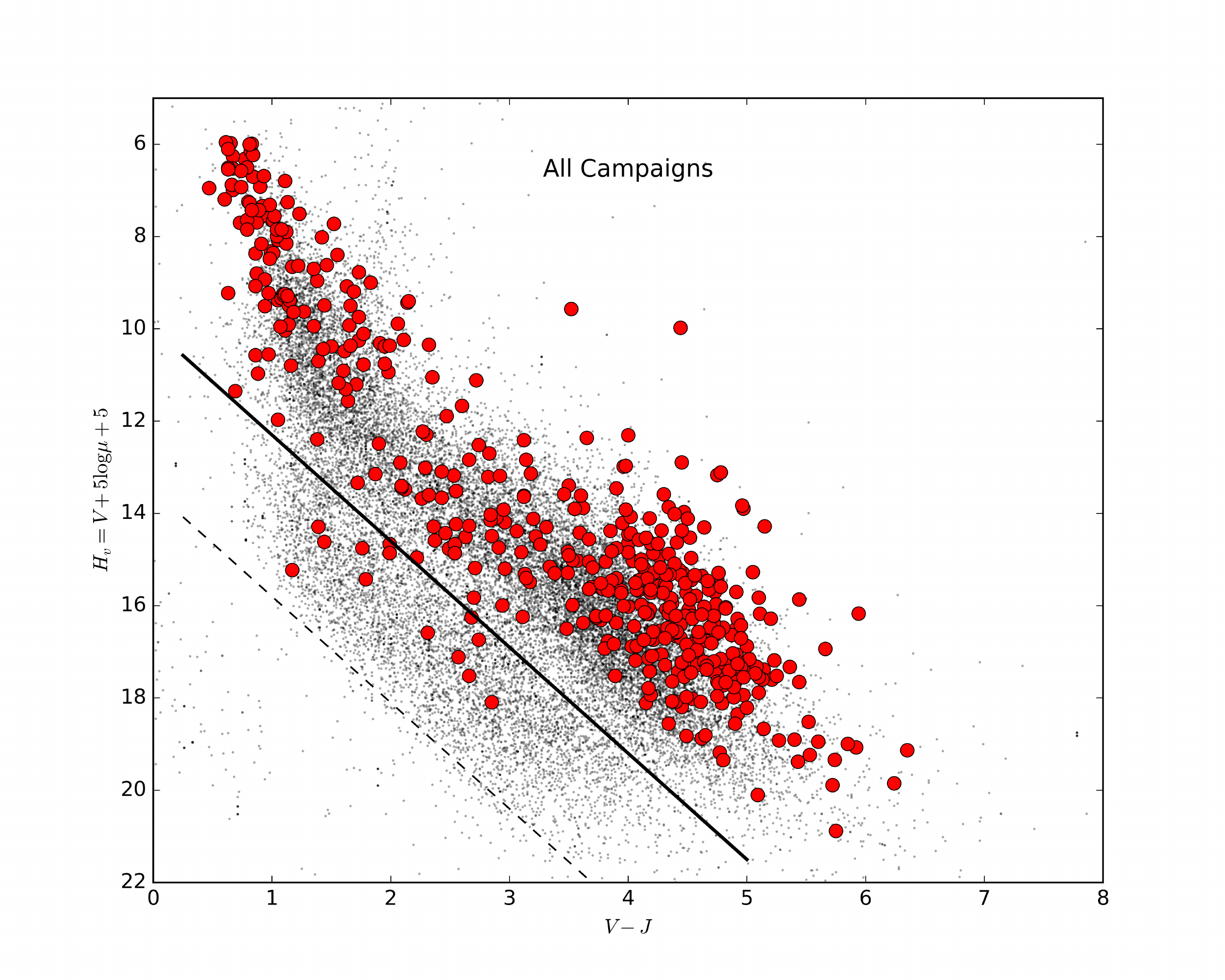}
\caption{Reduced proper motion diagram for all 27,382 high proper motion stars observed in the first nine \emph{K2} campaigns with light curves analyzed in the present study. The fast rotators are indicated by red circles. The solid black line represents an empirical division between the local Galactic disk (above line) and halo (below line) populations. The dashed black line represents an empirical division between the halo (above) and white dwarf (below) populations. The M dwarf fast rotators tend to be slightly elevated above the main locus of the Galactic disk population, which is a typical feature of nearby young stars. The rapidly rotating halo objects may be interacting binary systems, in which the rotation rates have been spun up by tidal interactions.}
\label{fig:rpm}
\end{figure}

We utilize the SUPERBLINK proper motions to produce a reduced proper motion (RPM) diagram of all the SUPERBLINK stars monitored by \emph{K2} (Figure \ref{fig:rpm}). The black solid line represents a ``by eye'' boundary separating the loci of the main-sequence disk and main-sequence halo stars, consistent with the metallicity analysis of \citet{SL07}. The black dashed line represents a ``by-eye'' division between the halo stars and the white dwarf population. The fast rotator candidates identified in our analysis are plotted as large red circles, with all other stars shown as small black dots. Strikingly, the vast majority of the fast rotator candidates are found within the locus of the disk stars, with only 13 candidates falling within the locus of halo stars. In addition, the M dwarf ($V-J>2.7$) fast rotators as a group are ``elevated'' (i.e. they have lower reduced proper motion values on average) within the locus of the disk stars compared to stars of similar colors. In the RPM diagram, this either means that these rapidly rotating M dwarfs tend to have brighter absolute magnitude at a given color or that they tend to have lower average transverse motions. A low average motion is consistent with youth because nearby star forming regions and young moving groups have relatively low transverse motions relative to the Sun. A higher absolute magnitude is also consistent with youth because very young ($<$ $100$ Myr) M dwarfs are not fully contracted onto the main sequence \citep{Bar96}. This feature means young M dwarfs are larger and more luminous for their color compared to the main sequence; therefore they are larger and more luminous compared to ZAMS stars of the same mass. Either one or a combination of both of these factors results in a smaller reduced proper motion value in young, local stars. That the fast rotators show the same expected trends suggest that these fast rotators are indeed part of the local young population.  \indent

Of the 27,382 SBK2 stars, 5,446 are halo objects based on their location in the RPM diagram, thus representing about $19\%$ of the sample. Of the 508 rapid rotators we identify, only 13 have reduced proper motions consistent with halo objects, which is less than $2\%$ of the fast rotators identified. The identification of any fast rotator among stars associated with the halo population is at first glance surprising, because these stars are mostly relatively old ($\gtrsim$ $10$ Gyr). One would expect that such old stars could be fast rotators only in the presence of a close companion, where the rotation rate is driven by tidal interactions. According to the data on nearby binaries and multiples presented in \citet{Deepak}, we would expect to find $\sim$ $0.4\%$ of stars in our sample to be interacting binary stars; this is based on the observation that only 2 binaries with orbital periods $<$ $5$ days are found among the 454 local stars that were surveyed in their study. For our subset of halo stars, we would thus expect $\sim$ $0.4\%$ of 5,446 stars to be close binaries and potentially interacting, which would be about 21 objects. The 13 candidate halo fast rotators we identify thus come reasonably close (within a factor 2) of the predicted number. The slightly lower than expected value may indicate that the binary fraction is different for the local halo population, compared to the nearby G stars surveyed in the Raghavan study. Alternatively, we may be missing some close binary stars that may not be sufficiently interacting or that have inclinations that make it difficult to detect rotation modulations. We are currently planning follow-up spectroscopic observations of some of these stars, as this small subset of halo stars is of special interest. Further analysis will be presented in an associated paper (Saylor et al., in preparation). \\ \indent
In any case, the much lower detection rate of fast rotators among the halo stars provides a convincing validation of our ACF analysis. Indeed extrinsic signals such as instrumental artifacts in the \emph{K2} light curves should be affecting all light curves equally regardless of what the disk/halo status of the star is. The very fact that the great majority of our candidate fast rotators are identified among ``disk'' stars shows that these modulations must be intrinsic signals, and thus most likely genuine modulations in those stars. In the worst case, if we assume that all 13 candidate fast rotators identified among the halo stars are false positives, this suggests a false positive detection rate of $0.05\%$ for all the stars in our SBK2 database. If this false positive rate were applied to the subset of disk stars, this means that approximately 65 of the fast rotators in the disk population might be false positives, or $12\%$ of all the candidate fast rotators identified in our study. \\ \indent
In addition, the fact that the halo stars are generally non-variable suggests that these stars, as a group, may be ideally suited for use as ``calibration'' sources to refine the reduction of K2 photometric data.

 \begin{figure}[H]
\centering
  \includegraphics[width=\linewidth,height=75mm]{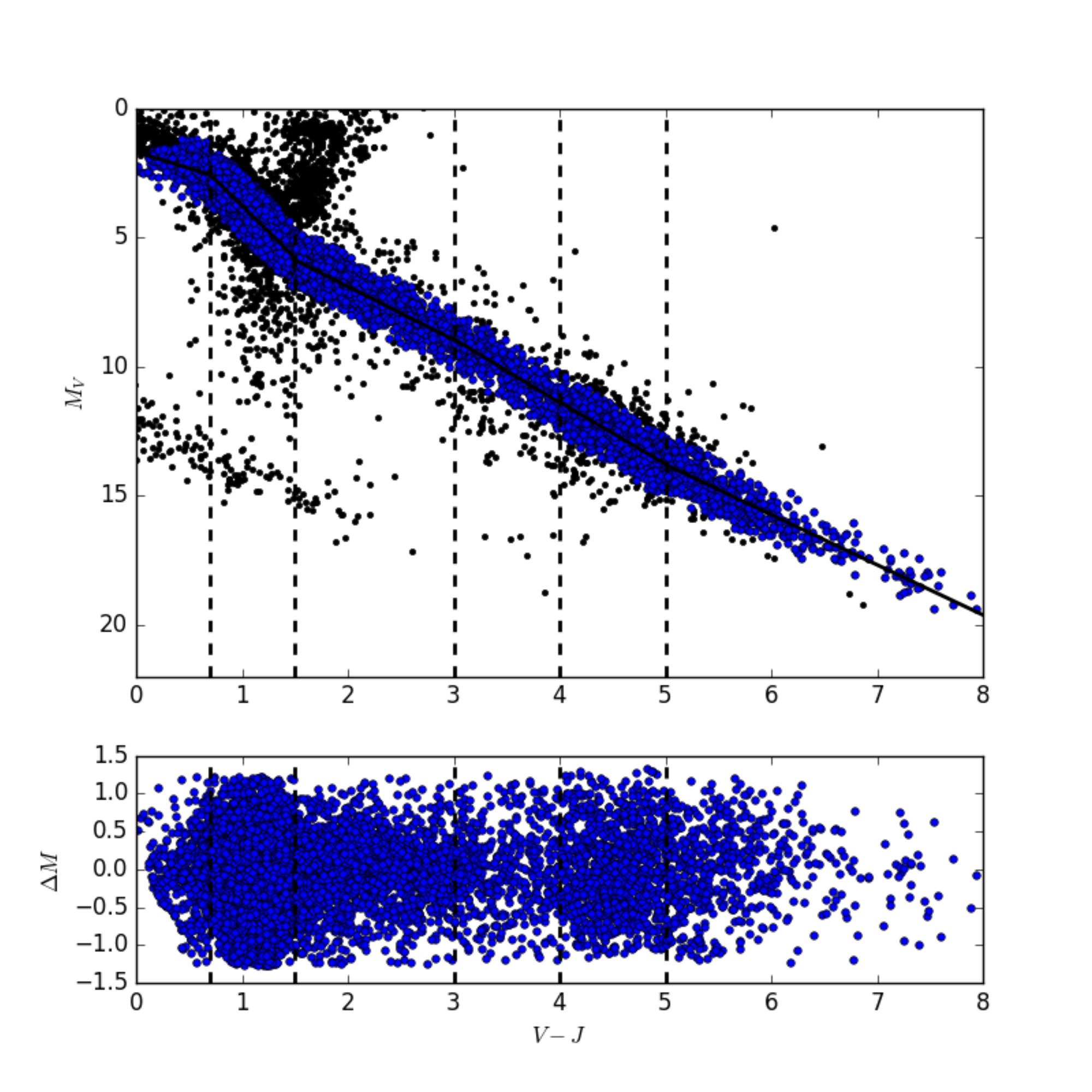}
\caption{Color-magnitude diagram used to derive photometric distances of SBK2 stars. All 11,809 calibration stars are plotted as black dots, and the stars that pass the cuts made to the initial polynomial fits are over plotted as blue dots. The bottom panel shows the difference between the derived magnitude and the actual magnitude of the calibration stars used to derive the fit. We find that the typical error on the photometric distances are $15\%$ of the distance, or $\pm 1$ in distance modulus.The equations of fit are listed in Section \ref{sec:projections}}
\label{fig:calib}
\end{figure}

\subsection{Velocity-Space Projections}
\label{sec:projections}
Most stars in the present sample do not have parallax measurements as of yet. Therefore, to evaluate the space motion of our stars, we combine proper motions with photometric distances derived from V magnitudes and V-J colors. We first identify 11,809 stars from the entire (all-sky) SUPERBLINK catalog with distances $d$ $<$ $100$ pc that currently do have measured trigonometric parallaxes (which includes parallaxes from the \emph{HIPPARCOS} and \emph{TGAS} catalogs) and calculate their absolute magnitudes $M_V$. We then determine the relationship between $M_V$ vs $V-J$ to use as a photometric distance calibration. Because the mathematical relationship between $M_V$ and $V-J$ is complex and varies with the $V-J$ color, we divide the stars into six $V-J$ bins, and fit a relationship for each one of the bins using a polynomial of order one. After a first pass, we eliminate any star which deviates from the fit by $\pm$1.5 magnitudes. We then fit a polynomial of order one to the remaining calibration stars. We repeat this process and total of three times. The best-fit linear equations for each of the magnitude bins are:

\begin{equation}
M_V =  1.26(V-J) + 1.67 \enspace  (V-J \leq0.7)
 \end{equation} \\
 \begin{equation}
M_V =  4.02(V-J) - 0.26  \enspace (0.7 < V-J\leq1.5) 
  \end{equation} \\
 \begin{equation}
M_V = 2.06(V-J) + 2.77 \enspace (1.5 < V-J\leq3.0)
  \end{equation} \\
 \begin{equation}
M_V = 2.47(V-J) + 1.51 \enspace (3.0 < V-J\leq4.0)
  \end{equation} \\
 \begin{equation}
M_V = 2.40(V-J) + 1.75 \enspace (4.0 < V-J\leq5.0) 
  \end{equation} \\
 \begin{equation}
 M_V = 1.94(V-J) + 4.08 \enspace (5.0 < V-J\leq6.0)
\end{equation}
 \label{eq:distance}
  
The top panel of Figure \ref{fig:calib} shows all 11,809 calibration stars as black dots, the stars that pass the cuts over plotted as blue dots, and the derived fit. The bottom panel shows the difference between the derived magnitude and the actual magnitude of the calibration stars that passed both cuts to the initial polynomial fits. We find that the typical error on the photometric distances,  $d_{phot}$, are $15\%$ of the distance, or $\pm$ 1 in distance modulus. \indent

\begin{figure*}
\centering
  \includegraphics[scale=0.5]{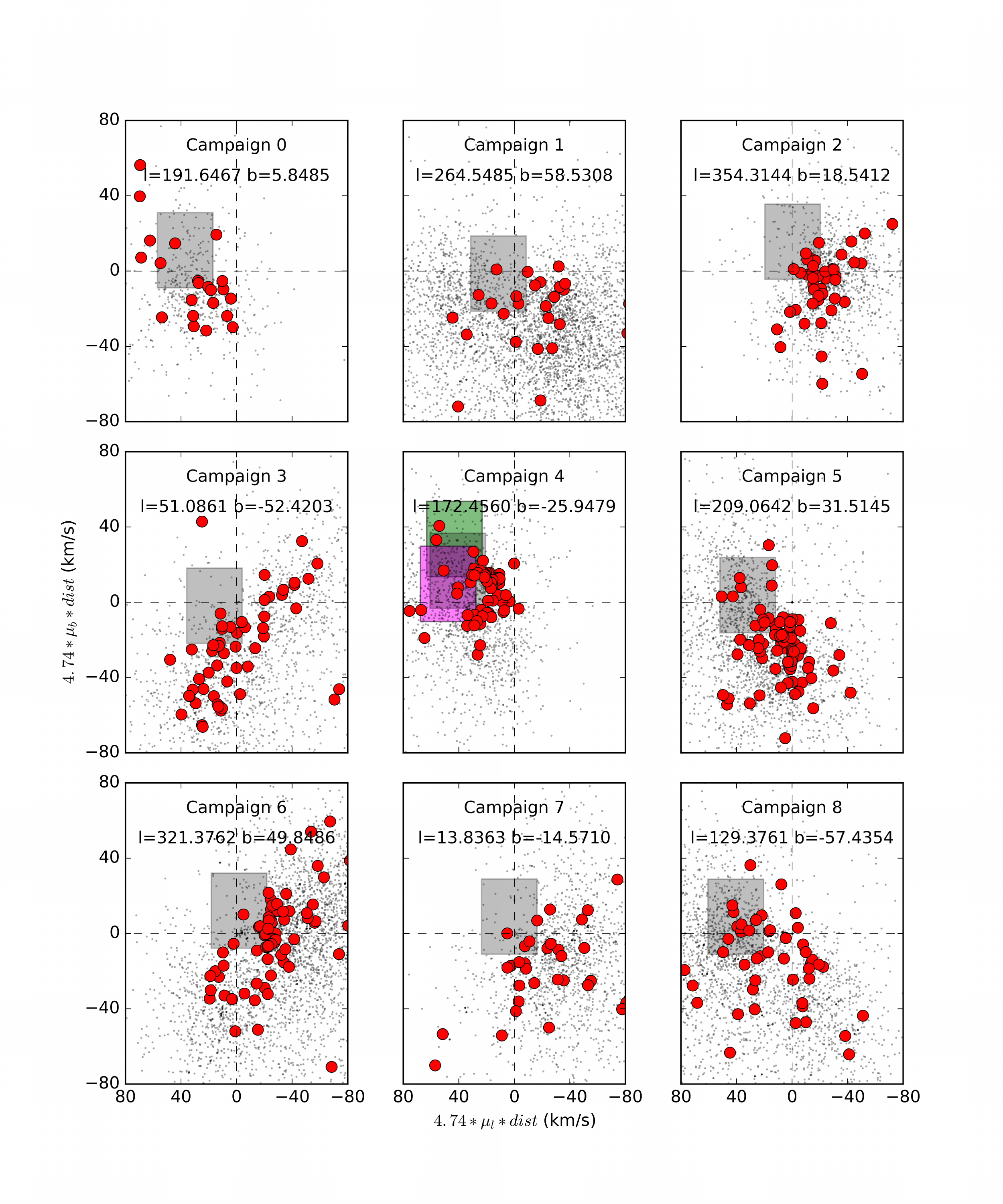}
\caption{Galactic velocity projection of all 27,382 SBK2 stars analyzed in this study. The black points are all SBK2 targets and the red circles are the subset of rapid rotators. The axes plot $4.74$ $\mu$ $d_{phot}$ which is the total proper motion multiplied by the photometric distance in units of km s$^{-1}$. The black box in each subfigure represents a generic characterization of the NYMG velocities. Field 4 specifically targeted the Hyades and Pleiades which are represented by the green and magenta boxes, respectively. Surprisingly, the fast rotators we identify do not appear to have velocities consistent with known NYMGs. The calculated photometric distances used in these diagrams assume that the star is settled on the main sequence. However, figure \ref{fig:rpm} indicates that these targets are elevated above the main sequence. Therefore, our distances are underestimated and our velocities are overestimated causing them to fall outside the boxes. We note clumps of fast rotators in campaigns 2, 4, and 5 which may be indicative of NYMGs. Further study of these star's distances and radial velocities is needed to determine group membership.}
\label{fig:velocities}
\end{figure*}

\begin{figure}[]
\centering
\subfigure{
  \includegraphics[scale=0.4]{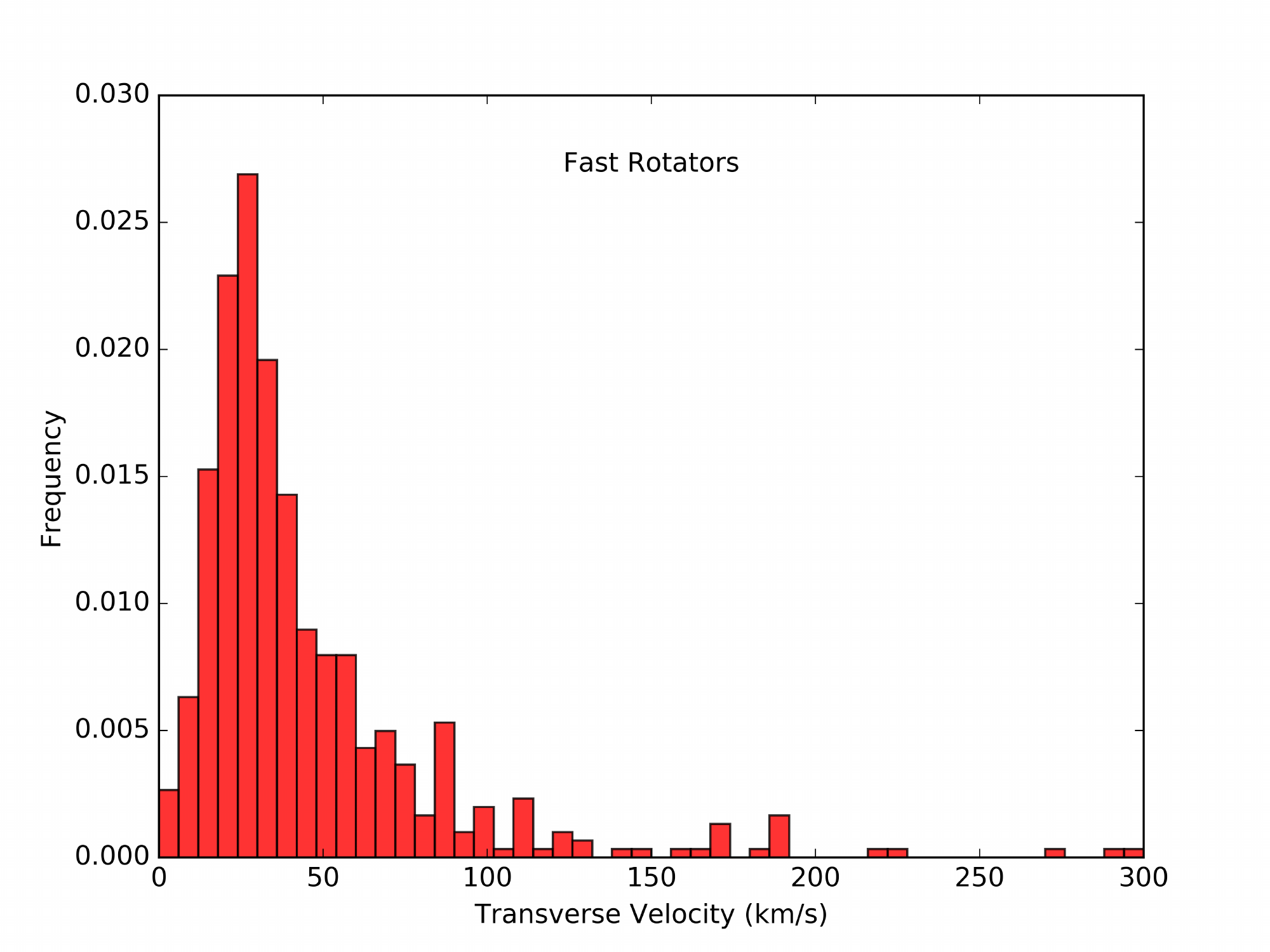}
}
\subfigure{
  \includegraphics[scale=0.4]{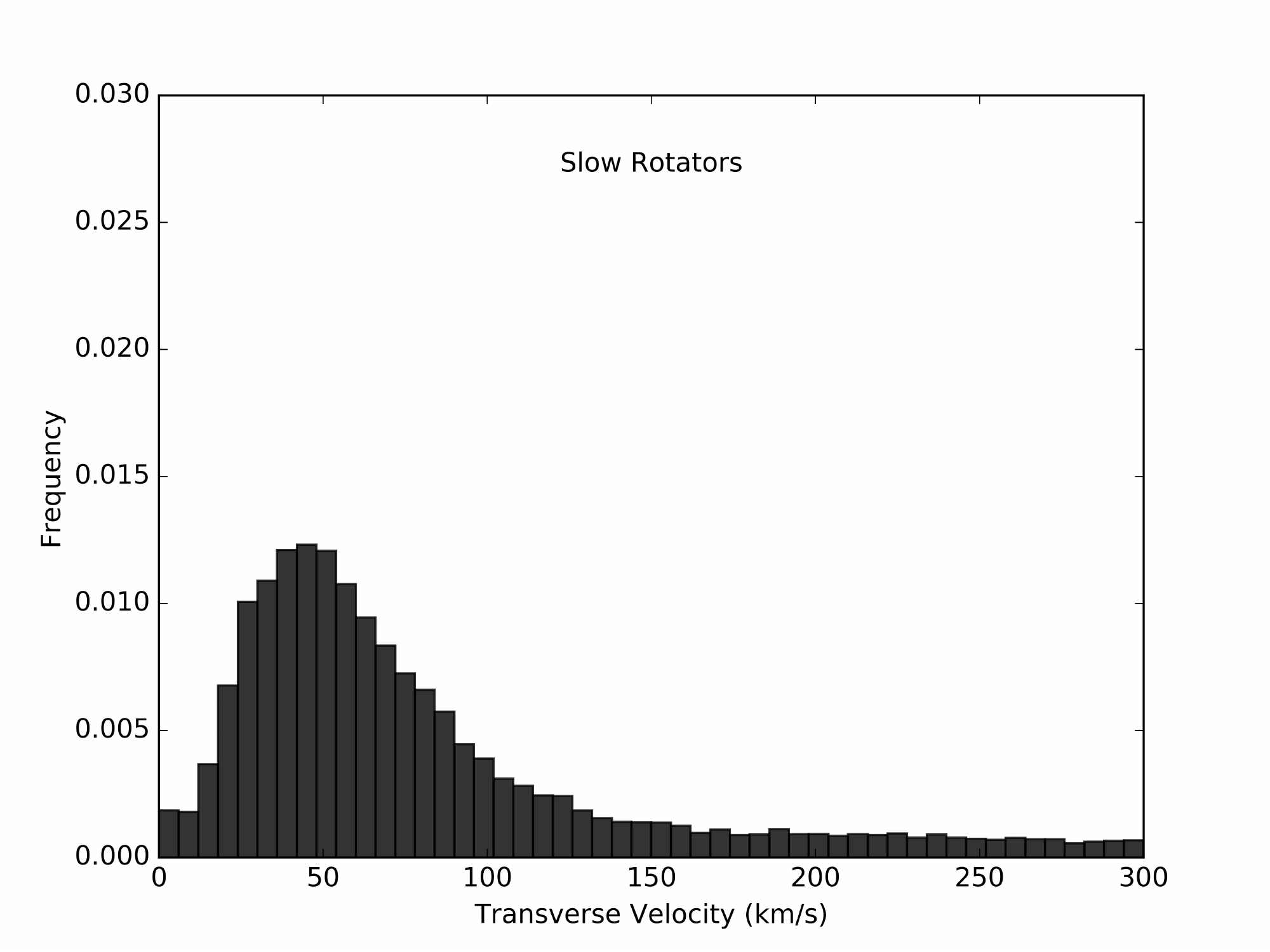}
  }
\caption{Frequency histograms of 2D velocity projections for the fast rotators (top) and all SBK2 stars (bottom). There are more fast rotators at low velocity and their overall distribution peaks at a lower velocity compared to the general population of stars, see Section \ref{sec:projections}. A two-sided KS test of these distributions produces a KS statistic of 0.37 and a p-value of $10^{-61}$ indicating that they are drawn from different populations.}
\label{fig:velhist}
\end{figure}

We multiply the photometric distances by the total proper motion, $\mu$, of each star to obtain a transverse velocity, from $v_T$ $=$ $4.74$ $\mu$ $d_{phot}$. Figure \ref{fig:velhist} shows a frequency histogram of those velocity space projections. The rapid rotators are shown on the top panel in red while all SBK2 stars are shown on the bottom panel in black. The rapid rotators appear to be more abundant at lower velocities and their distribution peaks at a lower velocity than the SBK2 stars as a whole. In fact, the median transverse motion of the fast rotators is 33 km s$^{-1}$, with a mean absolute deviation (MAD) of 19 km s$^{-1}$. For all the other stars, the median transverse motion is 60 km s$^{-1}$, with a MAD of 38 km s$^{-1}$, which implies that the fast rotators peak at a transverse velocity of about half that of the slow rotators. We believe this indicates that the population of rapid rotators is fundamentally different from the general population of stars, i.e. young thin disk stars vs. older thick disk $\&$ halo stars. A two-sided KS-test returns a KS statistic of 0.37 and a p-value of $10^{-61}$. These values strongly indicate that the motions of the fast rotators are indeed different from the general population of SBK2 stars. \\ \indent
We further combine these photometric distances with the measured proper motions to create kinematics plots. Although we do not have radial velocity data, the combination of proper motion vectors and distances are sufficient to create 2D projections of the space motions in the plane of the sky. We plot these 2D velocity projections in Figure \ref{fig:velocities} with individual plots for each one of the \emph{K2} campaign fields, since each field has a different plane of projection. The axes plot the projected velocity vectors ($v_l$,$v_b$) = $(4.74$ $\mu_l$ $d_{phot}$, $4.74$ $\mu_b$ $d_{phot})$, where $\mu_l$ and $\mu_b$ are the components of proper motion in Galactic coordinates, which are multiplied by the photometric distances derived from the procedure described above. The velocity components are given in units of km s$^{-1}$. The black box in each panel shows the approximate location in velocity space of most known nearby, young moving groups (NYMGs), using the generic velocities: U=-10.5 km s$^{-1}$, V=-19.5 km s$^{-1}$, W=-7.5 km s$^{-1}$.  The field of Campaign 4 targeted the Hyades and Pleiades; the projected velocity of those two clusters are represented by green and magenta box, respectively. We use the following velocities for the Hyades: U=-40 km s$^{-1}$, V=-18 km s$^{-1}$, W=-3 km s$^{-1}$, and for the Pleiades: U=-7 km s$^{-1}$, V=-27 km s$^{-1}$, W=-13 km s$^{-1}$. \\ \indent
A surprising result is the fact that most of the fast rotators we identify do not generally fall within these boxes, and thus do not appear to have kinematics consistent with the known NYMGs. While it is clear that the stars have low projected velocities consistent with the thin disk population, see Figure \ref{fig:velhist}, the velocity space distribution varies considerably between fields. Some fields show considerable spread of their fast rotators (e.g. campaign fields 3, 7, 8), while others show evidence of significant levels of clumping (fields 4, 5, 6). If it is true that young stars form in clusters or loose associations, then the more scattered objects in velocity space might be stars that have been ejected from their parent cluster/association, while the objects that form clumps in velocity space may be the still co-moving remnants of these groups. Ejected objects would be scattered throughout space and show up to some level in all \emph{K2} fields, while the more compact remnants (NYMGs) would be apparent only in some fields. An alternative explanation could be due to the fact that the calculated photometric distances used in these diagrams assume that the star is settled on the main sequence, which would result in significant uncertainty in the photometric distance for M dwarfs that are very young and still contracting. Indeed, Figure \ref{fig:rpm} does suggest that some of these stars are elevated above the main sequence. Therefore, our distances for the fast rotators are potentially underestimated, which would mean that our velocities are overestimated. A revised distance calibration might move our targets more in line with the velocities of the NYMG. \indent

Tables 3 and 4 list the median and mean average deviation (MAD) of the 2D velocity space projections of the fast and slow rotators, respectively, for each \emph{K2} field analyzed in this work. The average ratio of the MAD for the fast rotators compared to the entire SBK2 sample is 0.65 in the Galactic latitude direction and 0.68 in the Galactic longitude direction. This fraction implies that the fast rotator distribution of velocities is $40\%$ smaller than the general population of targets we analyzed despite the fact we are potentially overestimating their transverse velocities. This tighter distribution of velocities is consistent with the young, thin disk. We therefore conclude the fast rotators are drawn from a younger population.

\section{Stars with UV Excess}
\label{sec:uvexcess}
\begin{figure}[]
\centering
 \includegraphics[width=\linewidth,height=75mm]{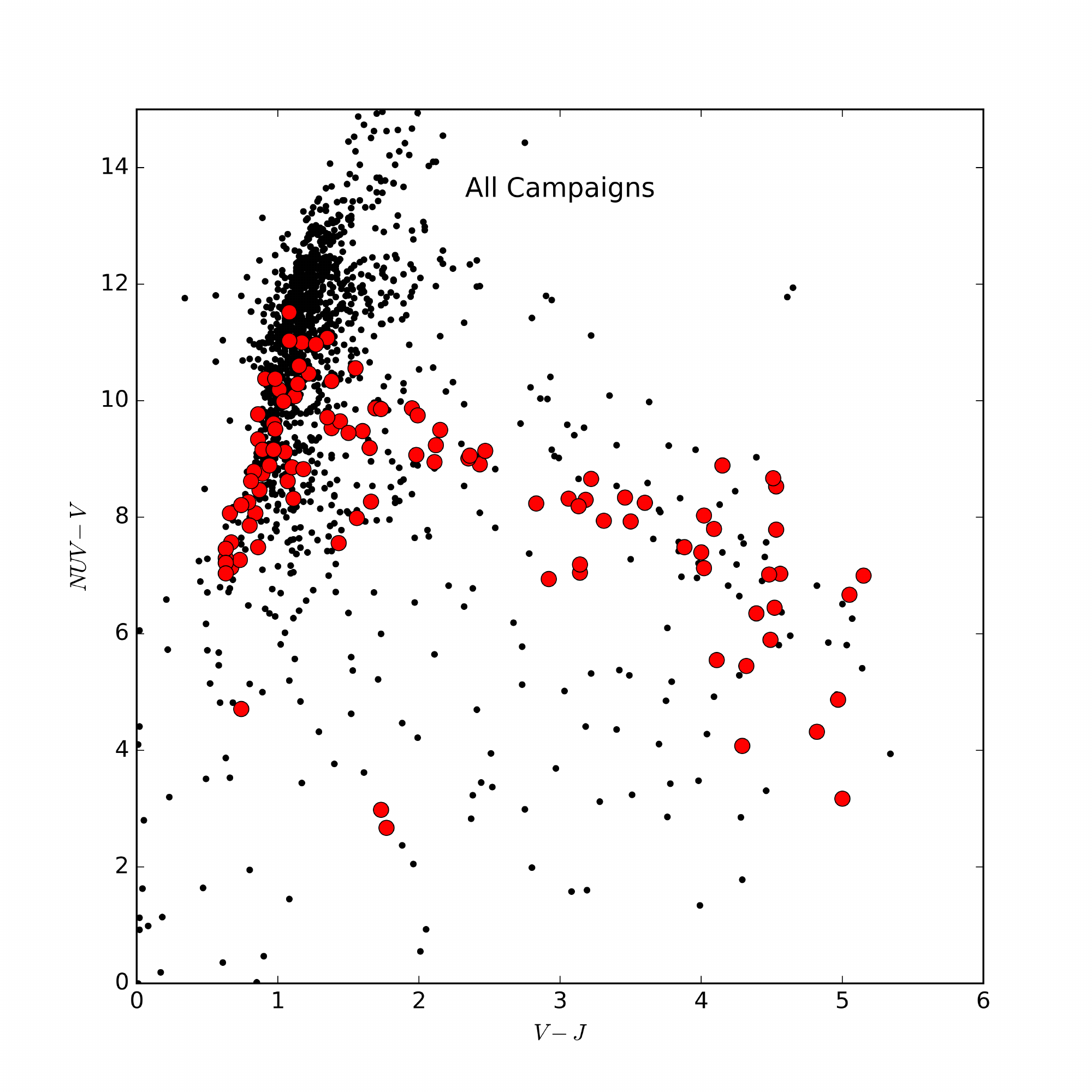}
\caption{NUV-V vs V-J for SBK2 targets. Black points are all SBK2 stars while red circles are the rapid rotators with \emph{GALEX} magnitudes. We find that the M dwarfs identified as fast rotators have a $20\%$ chance of showing UV excess, see Table \ref{tbl:galexobs}.}
\label{fig:colorcolor}
\end{figure}

We cross-matched all SBK2 sources with the GALEX catalog of UV sources to investigate whether fast rotators would also normally be detected as chromospherically active stars from excess  UV emission. Figure \ref{fig:colorcolor} is a color-color diagram for all SBK2 stars with a counterpart in the \emph{GALEX} source catalog. We utilize \emph{GALEX} NUV fluxes (when available) as well as V and J magnitudes. The black points are all SBK2 targets with \emph{GALEX} counterparts; the red circles are the fast rotators identified in this work. We define a star to be an M dwarf is it has $V-J$ $>$ $2.7$, following \citet{SL13}. We further define an M dwarf to be ``active'' if it has $NUV-V$ $<$ $8$, indicative of a significant UV excess. Using these criteria, we identify a total of 124 ``active M dwarfs'' in the SBK2 sample. Of these, we find that 25 are also identified as fast rotators, which represents $\sim20\%$ of the ``active M dwarf'' subset. By comparison, there are 15,745 SBK2 stars that have M dwarf colors, and of these only 328 are fast rotators, or a fast rotator rate of just $2\%$. These statistics indicate that the M dwarf fast rotators selected from the \emph{K2} light curves are significantly more likely to have UV excess, which supports the connection between fast rotation and chromospheric activity. We additionally note that for the stars which are not M dwarfs, i.e. $V-J$ $<$ $2.7$, there is no evidence that the fast rotators have significant NUV excess over the other stars. This evidence indicates that NUV excess is not an entirely reliable age diagnostic for higher mass fast rotators. \indent

\begin{deluxetable}{ccccc}
\tablenum{3}
\tablecaption{Velocity Distribution of Fast Rotators}
\tablewidth{0pt}
\tablehead{
\colhead{Field} & \colhead{$<v_l>_{med.}$ } & \colhead{$MAD_{v_l}$ } & \colhead{ $<v_b>_{med.}$} & \colhead{$MAD_{v_b}$}  \\
& (km s$^{-1}$) & (km s$^{-1}$) & (km s$^{-1}$) & (km s$^{-1}$)}
\startdata
0 & 27 & 26 & -9 & 20 \\
1 & -16 & 23 & -17 & 15 \\
2 & -20 & 13 & -3 & 14 \\
3 & 9 & 22 & -24 & 31 \\
4 & 21 & 1 & 8 & 11 \\
5 & 6 & 15 & -24 & 15 \\
6 & -35 & 18 & -2 & 21 \\
7 & -34 & 40 & -25 & 17 \\
8 & 35 & 63 & -24 & 32 
\enddata
\label{tbl:fastvelocity}
\end{deluxetable}

\begin{deluxetable}{ccccc}
\tablenum{4}
\tablecaption{Velocity Distribution of all SBK2 Targets}
\tablewidth{0pt}
\tablehead{
\colhead{Field} & \colhead{$<v_l>_{med.}$ } & \colhead{$MAD_{v_l}$ } & \colhead{ $<v_b>_{med.}$} & \colhead{$MAD_{v_b}$}  \\
& (km s$^{-1}$) & (km s$^{-1}$) & (km s$^{-1}$) & (km s$^{-1}$)}
\startdata
0 & 31 & 22 & -12 & 19 \\
1 & -21 & 71 & -42 & 43 \\
2 & -28 & 23 & -4 & 21 \\
3 & -11 & 47 & -34 & 40 \\
4 & 41 & 39 & -2 & 34 \\
5 & 36 & 53 & -38 & 41 \\
6 & -51 & 63 & -14 & 51 \\
7 & -40 & 34 & -15 & 31 \\
8 & 8 & 47 & -31 & 49 
\enddata
\label{tbl:slowvelocity}
\end{deluxetable}

Approximately $95\%$ of the rapid rotators we identify do not fit the criterion for UV excess. Part of this is due to the patchy coverage of GALEX. In particular, \emph{GALEX} did not observe many regions close to the Galactic plane due to high field density. Of the fields we analyzed, C0 and C7 fall directly on the Galactic plane. Campaigns C2 and C4 lie near the Galactic plane, while C1, C3, C5, C6, and C8 are well within the coverage of the \emph{GALEX} survey. Table 5 presents a breakdown of the number of SBK2 M dwarfs with UV excess per field and the number of those stars which are identified as fast rotators. \\ \indent
Table 5 indicates that most of the fast rotator M dwarfs we identify are not in the low Galactic latitude fields. Of the fields that are well covered by GALEX, we still find that only $\sim$ $20\%$ of the SBK2 M dwarfs show evidence of UV excess from GALEX data. Therefore, it would appear likely that many of the rapid rotators in our sample do have some level of UV excess from chromospheric activity, but \emph{GALEX} was unable to detect these targets, possibly due to the intrinsic faintness of those M dwarfs. We believe that this is strong evidence that the search for rapid rotators in fields observed with \emph{Kepler} and \emph{K2} can identify young stars that lack a UV detection, which can significantly expand the search for nearby young M dwarfs. \indent

An interesting pattern that emerges in Figure \ref{fig:colorcolor} is the fact that most of the M dwarf fast rotators have $NUV-V$ $>$ $6$. That is, the majority of M dwarfs with larger UV excess ($NUV-V$ $<$ $6$) are not identified as fast rotators in our study. This is consistent with the observation that most M dwarfs with very large NUV excess are actually low-mass stars with a white dwarf companion (Skinner et al. 2017 - in press). The white dwarf is responsible for the large NUV excess, but these systems are of course relatively older, which means one would not expect the M dwarf component to be a fast rotator, unless the rotation of the M dwarf is driven by tidal forces in a tight system. \\ \indent
The technique we describe in this paper could prove especially useful once data from the upcoming \emph{TESS} mission becomes available. These data will make it possible to expand the search for young M dwarfs over much of the sky, especially if most of the known nearby M dwarfs are being targeted by \emph{TESS}. 

\setlength{\tabcolsep}{0.3in}
\begin{deluxetable}{ccc}
\tablenum{5}
\tablecaption{Number of M Dwarf Fast Rotators and Fast Rotators with Detected GALEX NUV Emission}
\tablewidth{0pt}
\tablehead{
\colhead{Field} & \colhead{Fast Rotation} & \colhead{Fast Rotation} \\
& & \colhead{ $+$NUV} }
\startdata
0 & 1 & 0 \\
1 & 25 & 0 \\
2 & 1 & 0 \\
3 & 8 & 2 \\
4 & 2 & 1 \\
5 & 42 & 9 \\
6 & 12 & 5 \\
7 & 0 & 0 \\
8 & 33 & 8 
\enddata
\label{tbl:galexobs}
\end{deluxetable} 

\section{Conclusions}
\label{sec:conclusions}
We have searched the first nine \emph{K2} campaigns for SUPERBLINK high proper motion stars with rotation periods $P_{rot}<4$ days. According to gyrochronolgy, this should correspond to an age of $\lesssim 150$ Myr. We developed a data analysis pipeline that makes use of an auto-correlation function (ACF) combined with a Fast Fourier transform (FTT) to identify modulation in the \emph{K2} light curves, eliminate obvious instrumental signals, and estimate the timescale of the intrinsic modulation. In most cases the variability appears to be consistent with starspots on a fast-rotating star, so the characteristic time-scale modulations are assumed to be estimates of the stellar rotation period.This combined ACF and FFT algorithm identifies 508 probable fast rotators in the first nine campaign fields of the \emph{K2} mission (C0 to C8). \\ \indent
Based on their location in a reduced proper motion diagram, we determine that nearly all the fast rotators are consistent with the thin disk population. However, we also identify 13 rapid rotators with colors and/or kinematics consistent with the old halo population. We believe that these stars may be tidally interacting binary systems. The second paper in this series will discuss this hypothesis. \\ \indent
More generally, the reduced proper motion diagram shows that the fast rotators tend to have relatively low reduced proper motions at a given color compared to normal field stars. This either indicates that the stars have low transverse motions or that they are elevated above the ZAMS (i.e. overluminous) or both. In either case, this trend suggests that they are part of a younger field population. A frequency distribution of the transverse motions, estimated using photometric distances, clearly shows that there are more fast rotators at lower velocities than the general population of SBK2 targets and that the distribution peaks at a lower transverse velocity. Two-dimensional plots of the transverse motion vectors in the plane of the sky further shows significant differences in the kinematics of fast-rotators in the different \emph{K2} campaign fields, along with indications of significant velocity-space clustering in campaigns 2, 4, and 5. This is consistent with the idea that young field stars are part of various collections of young moving groups, resulting from the slow dispersion of loosely bound stellar associations. Further study of these star's distances and radial velocities is needed to determine group membership.\\ \indent
A color-color plot of NUV-optical-IR magnitudes ($NUV-V$,$V-J$) indicates that a significant fraction (20$\%$) of the fast rotators we identify also have NUV excess detected in the \emph{GALEX} survey. This indication of chromospheric activity further confirms that these stars are young. However, we find that in \emph{K2} fields with good \emph{GALEX} coverage, $\sim$ $95\%$ of the fast-rotator M dwarf candidates are not detected in \emph{GALEX}; this suggests that the absence of a detected UV excess is not sufficient to rule out youth in a local field M dwarf, and that starspot modulation may be more efficient in identifying nearby young, low-mass stars. We therefore conclude that the \emph{K2} data, along with upcoming data from the NASA \emph{TESS} mission, can be used most effectively as a way to identify the population of young M dwarfs near the Sun.  \\ \indent
The \emph{TESS} mission should obtain light curves for most nearby M dwarfs unlike \emph{Kepler} and \emph{K2} which do not cover much of the sky. The technique we describe in this paper could be used to search for young M dwarfs \emph{independent} of NUV coverage. 

\section*{Acknowledgements}
This material is based on work supported by the National Science Foundation under grant No. AST 09-08419, the National Aeronautics and Space Administration under Grants No. NNX15AV65G, NNX16AI63G, and 
NNX16AI62G issued through the SMD/Astrophysics Division as part of the  K2 Guest Observer Program. 

\appendix 
\section{List of SUPERBLINK stars in Kepler campaigns C0-C8}
The complete list of 27,382 SBK2 targets can be found in Table 6. The first 20 lines of the table are shown in the print version; the complete list is available in the electronic version of the paper. \\ \indent
Columns 1 and 2 give the identification number of the star in the SUPERBLINK and EPIC catalogs, respectively. Columns 3 and 4 list the J2000 right ascension and declination, respectively. Columns 5 and 6 list the SUPERBLINK catalog proper motion in the direction of right ascension and declination, respectively, both in units of seconds of arc per year. Column 7 lists NUV magnitude from \emph{GALEX}, when available. Column 8 lists V magnitudes from SUPERBLINK. Column 9 lists V-J color using 2MASS infrared magnitudes, J. Column 10 is the \emph{K2} field corresponding to the target. 

\setlength{\tabcolsep}{0.15in}
\begin{deluxetable*}{lrrrrrrrrr}
\tablenum{6}
\tabletypesize{\small}
\tablecaption{SUPERBLINK Stars Observed in \emph{K2} Campaigns 0-8}
\tablewidth{0pt}
\tablehead{
\colhead{SB} & \colhead{EPIC} & \colhead{$\alpha$} & \colhead{$\delta$ }  & \colhead{$pm_{\alpha}$}   & \colhead{$pm_{\delta}$}  & \colhead{NUV} & \colhead{V}  & \colhead{V-J} &  \colhead{Field}   \\
& & \colhead{(J2000)} & \colhead{(J2000)} & \colhead{(mas/yr)} & \colhead{(mas/yr)} }
\startdata
PM I00341$+$0436		&	220394397	&	8.539574	&	4.601908	&	0.0464	&	-0.0254	&	99.99	&	17.59	&	3.89	&	8	\\
PM I00348$+$0433		&	220392429	&	8.712526	&	4.563131	&	0.0604	&	-0.0018	&	99.99	&	12.14	&	1.66	&	8	\\
PM I00349$+$0422N		&	220383386	&	8.739684	&	4.381468	&	0.1073	&	-0.1734	&	22.38	&	8.97	&	1.42	&	8	\\
PM I00351$+$0406		&	220370126	&	8.780957	&	4.108527	&	0.0016	&	-0.0665	&	99.99	&	17.14	&	4.12	&	8	\\
PM I00351$+$0358		&	220363792	&	8.785554	&	3.978688	&	0.1203	&	0.0205	&	99.99	&	16.73	&	3.90	&	8	\\
PM I00352$+$0356		&	220361750	&	8.810661	&	3.937430	&	0.0798	&	0.0370	&	99.99	&	11.73	&	1.70	&	8	\\
PM I00352$+$0400		&	220365159	&	8.818637	&	4.007222	&	-0.0292	&	-0.0424	&	99.99	&	13.46	&	1.67	&	8	\\
PM I00353$+$0442		&	220399282	&	8.836354	&	4.701310	&	-0.0324	&	-0.0224	&	99.99	&	15.00	&	1.56	&	8	\\
PM I00355$+$0432		&	220391351	&	8.876101	&	4.540965	&	0.0862	&	-0.0393	&	99.99	&	19.62	&	3.25	&	8	\\
PM I00356$+$0438		&	220396177	&	8.902689	&	4.639234	&	0.0368	&	-0.0082	&	99.99	&	13.00	&	1.47	&	8	\\
PM I00357$+$0357		&	220362983	&	8.929224	&	3.963239	&	-0.0322	&	-0.1250	&	18.43	&	18.88	&	0.02	&	8	\\
PM I00357$+$0445		&	220402286	&	8.939465	&	4.760565	&	0.0368	&	-0.0136	&	99.99	&	15.59	&	3.00	&	8	\\
PM I00357$+$0439		&	220396738	&	8.941160	&	4.650753	&	0.1505	&	0.0357	&	99.99	&	20.40	&	3.67	&	8	\\
PM I00357$+$0440		&	220397535	&	8.942800	&	4.667059	&	0.1691	&	0.1022	&	99.99	&	19.39	&	4.83	&	8	\\
PM I00358$+$0354		&	220360455	&	8.954550	&	3.910664	&	0.0486	&	0.0076	&	99.99	&	16.52	&	2.18	&	8	\\
PM I00359$+$0352		&	220358904	&	8.983678	&	3.878616	&	0.1117	&	-0.0655	&	99.99	&	16.40	&	3.92	&	8	\\
PM I00359$+$0442		&	220399472	&	8.999578	&	4.704960	&	0.0705	&	0.0224	&	99.99	&	14.95	&	4.21	&	8	\\
PM I00361$+$0457		&	220412119	&	9.037066	&	4.952855	&	-0.0379	&	-0.0389	&	99.99	&	13.72	&	2.11	&	8	\\
PM I00363$+$0349		&	220355964	&	9.076753	&	3.820655	&	-0.0270	&	-0.0714	&	99.99	&	14.93	&	2.53	&	8	\\
PM I00363$+$0342		&	220350324	&	9.096736	&	3.704963	&	0.0557	&	-0.0044	&	99.99	&	11.99	&	1.71	&	8	
\enddata
\tablecomments{The full table is available in the electronic version of the paper.}
\label{tbl:slowrot}
\end{deluxetable*}


\end{document}